\documentclass[fleqn,usenatbib]{mnras}

\usepackage{newtxtext,newtxmath}

\usepackage[T1]{fontenc}

\DeclareRobustCommand{\VAN}[3]{#2}
\let\VANthebibliography\thebibliography
\def\thebibliography{\DeclareRobustCommand{\VAN}[3]{##3}\VANthebibliography}

\usepackage{graphicx}	
\usepackage{amsmath}	
\usepackage[percent]{overpic}
\usepackage{gensymb}

\title[A multi-wavelength view of our local ISM]{A three-dimensional, multi-wavelength view and time-dependent analysis of the Milky Way's local ionized gas}

\author[L. McCallum et al.]{Lewis McCallum,$^{1}$\thanks{E-mail: lm261@st-andrews.ac.uk}
Kenneth Wood,$^{1}$
Robert Benjamin$^{2}$, Dhanesh Krishnarao$^{3}$, Anna F. McLeod$^{4,5}$
\\
$^{1}$ School of Physics and Astronomy, University of St Andrews, North Haugh, St Andrews, KY16 9SS, UK\\
$^{2}$ Department of Physics, University of Wisconsin-Whitewater, Whitewater, WI 53190, USA\\
$^{3}$ Department of Physics, Colorado College, Colorado Springs, CO 80903, USA\\
$^{4}$ Centre for Extragalactic Astronomy, Department of Physics, Durham University, South Road, Durham DH1 3LE, UK \\
$^{5}$ Institute for Computational Cosmology, Department of Physics, University of Durham, South Road, Durham DH1 3LE, UK\\
}

\date{Accepted XXX. Received YYY; in original form ZZZ}

\pubyear{\the\year{}}

\begin{document}
\label{firstpage}
\pagerange{\pageref{firstpage}--\pageref{lastpage}}
\maketitle

\begin{abstract}

This work is the continuation of a series attempting to characterize the local warm ionized medium through both static and time dependent simulations. We build upon our three dimensional, observationally-derived simulation of the local photoionized interstellar medium - based on static photoionization simulations constrained by 3D dust maps - to include metals required to predict collisionally excited optical and infrared emission lines, providing the first all-sky prediction of a series of lines including [S{\sc ii}] 6716{\AA}, [N{\sc ii}] 6584{\AA} and [O{\sc iii}] 5007{\AA}. While these predictions only include O-star photoionization under ionization equilibrium, we also carry out a suite of radiation-hydrodynamics simulations including time-dependent metal ionization and the effects of supernova feedback to highlight missing features in our predicted skies. We use the simulations to estimate the very local (1~$\rm kpc^{2}$) Galactic star formation rate, finding a rate of 370~$\rm M_{\odot}~Myr^{-1}~kpc^{-2}$ provides the best match between the observationally-derived and ab-initio simulations. This is approximately a factor of four lower than previous estimates for the star formation rate required to support an observed layer of high-altitude diffuse ionized gas, possibly suggesting a `bursty' star formation history in the region surrounding the Sun. We also investigate the effects of O-star environments on their ability to ionize large volumes of diffuse ionized gas, and find it is likely ionized by a small number of luminous O-stars located in regions where the leakage of their Lyman continuum photons can produce the vast volumes of ionized gas observed in the midplane and at high galactic altitudes. 

\end{abstract}




\section{Introduction}

There are two principal methods to determine the record of recent (within the last 80 Myr) star formation in our solar neighborhood. First, one can study the nearby interstellar medium (ISM) for nearby interstellar shells, filaments, clouds, and outflows. These structures provide evidence of the energy input back into the Solar environs from recent generations of massive stars, e.g. \citet{1987ARA&A..25..303C}, \citet{1998ApJ...498..689H}, \citet{2011ARA&A..49..237F}, \citet{2023ASPC..534...43Z}.
And second, one can identify and characterize young clusters and OB associations near the Sun, examine their ages, dynamics, and chemical abundances, and search for commonalities in their histories, e.g. \citet{1964ARA&A...2..213B}, \citet{1999ASIC..540..411B}, \citet{2023ASPC..534..129W}.    

For the first approach, advances in the three-dimensional mapping of dust \citep{2006A&A...453..635M,2014MNRAS.443.2907S,2014ApJ...789...15S,2015ApJ...810...25G,2019ApJ...887...93G,2019A&A...625A.135L,2022A&A...661A.147L,2019MNRAS.483.4277C,2022A&A...664A.174V,2020A&A...639A.138L,edenhofer23,2024A&A...692A.255R,2024A&A...685A..39M,2024MNRAS.532.3480D,2025arXiv250302657Z} constructed by obtaining extinction measurements \citep{2019A&A...628A..94A,zhang23} and parallaxes for millions of stars \citep{2016A&A...595A...1G} have dramatically improved our understanding of the topology of the interstellar medium out to distances of three kiloparsecs from the Sun \citep{2023ASPC..534...43Z}. These maps provide new details on numerous interstellar structures, including the ``Radcliffe Wave,'' a section of the Orion Arm \citep{1953ApJ...118..318M} just outside the Sun's orbit characterized by a kiloparsecs-long dust structure with both wavy morphology \citep{2020Natur.578..237A} and oscillatory kinematics \citep{2022A&A...660L..12T,2024Natur.628...62K}, and ``the Split,'' a kiloparsecs-long dust band just inside the solar orbit \citep{2019A&A...625A.135L,2019ApJ...887...93G} embedded with shells and star formation.

 \begin{figure*}
    \centering
    \includegraphics[width=1.0\textwidth]{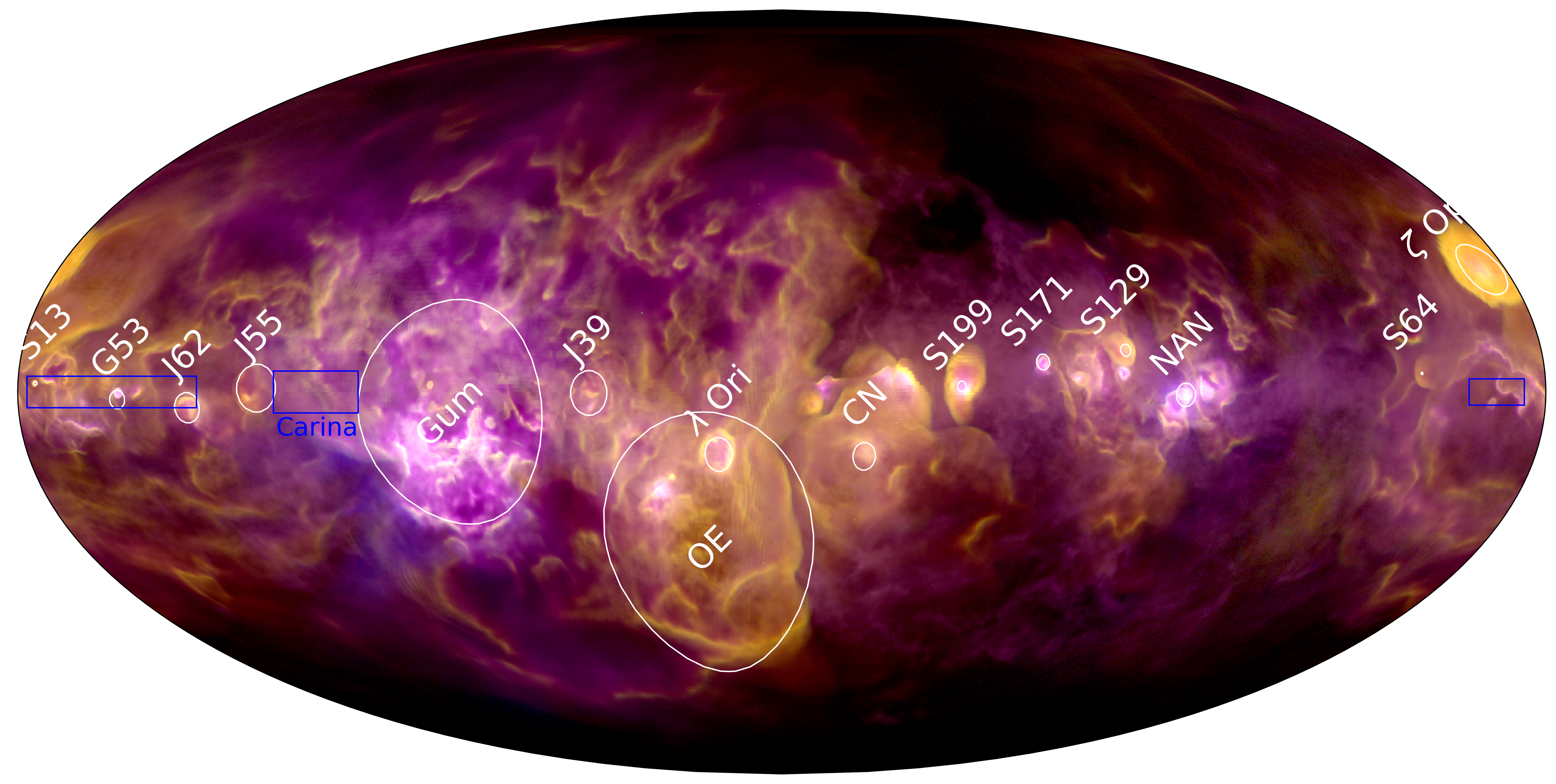}
    \caption{All-sky map of our model of the local ionized ISM, described in section~\ref{observationalmodel}, based on radiation transfer models incorporating the 3D dust map of \citet{edenhofer23} with the positions, luminosities and temperatures of 87 O stars in the solar neighborhood. In this three-color image, the H$\alpha$ emission is shown in red, [S{\sc ii}] 6716{\AA} in green and [O{\sc iii}] 5007{AA} in blue. Map projection is centered on $l=180\degree$, and the image is annotated with selected H{\sc ii} regions within 1.25 kpc as in \citet{mccallum25}, most prominently the Gum Nebula, the Orion-Eridanus (OE) superbubble, and the North American Nebula (NAN). The blue boxes show the direction of bright H{\sc ii} regions that lie beyond the 1.25 kpc simulation volume. }
    \label{linepanels}
\end{figure*}

There have also been major advances in characterizing the spatial distribution and trajectories of nearby massive stars and their associated young clusters as well as the history of star formation in the solar environs \citep{2011AJ....142..197C,2012ARA&A..50..531K,2016ApJ...831...73V,2018ApJ...864..136B,2019A&A...628A.123Z,2019MNRAS.487.1400C,2020NatAs...4..965R,2021MNRAS.504.2968P,2021A&A...651A.104P,2021A&A...645L...8X,2022MNRAS.516.4245W,2022MNRAS.509.2339N,2022A&A...664L..13S,2022ApJ...941..162E,2023A&A...678A..95S,2023A&A...669A..10Z,2025ApJ...980..216E,2025MNRAS.538.1367Q,2025A&A...696A..67S}. Of particular note is \citet{swiggum24}, who used the positions, velocities, and ages from a recently published catalog of stellar clusters from \citet{2023A&A...673A.114H}, supplemented by radial velocities from other programs and a catalog of Young Local Associations \citep{2018ApJ...856...23G}, to show that at least 57\% of 272 young ($<$70 Myr) clusters within one kiloparsec of the Sun originated in three distinct star-forming complexes. Each complex had an estimated gas mass of 1-2 million solar masses which started to form stars over 45 Myr ago, resulting in over 200 supernovae and three ``families'' of clusters: the $\alpha$Per family (82 clusters and young associations ), the M6 family (34 clusters), and the Collinder 135 family (39 clusters). A smaller, fourth family of eight clusters, the $\gamma$ Velorum family, was also identified.  

These two approaches have started to merge with individual stellar clusters and associations being plausibly identified as the energy source for nearby interstellar structures.\footnote{Although there is a long history of such efforts, c.f. \citet{1999ASIC..540..411B}, the precision of stellar positions and velocities now available with {\it Gaia} measurements make these asssociation significantly more secure.} The origin of the Local Bubble has been attributed to the Upper Centaurus-Lupus (UCL) and LCC (Lower Centaurus-Crux) clusters of the $\alpha$Per family \citep{2022Natur.601..334Z,swiggum24}, the origin of the Gum Nebula to the Vela OB2 association of the $\gamma$ Velorum family \citep{swiggum24,2025arXiv250412381G}, and the origin of the HI supershell GSH 238+00+09 \citep{1998ApJ...498..689H} to the clusters of the Collinder 135 family with some assistance from clusters of the M6 family. The linkage between some nearby bubbles/shells with local clusters is a bit less clear, e.g., the Orion-Eridanus superbubble \citep{2023ApJ...947...66F} and the Perseus-Taurus shell \citep{2021ApJ...919L...5B}, but some connection is probable.  
  
In order to associate shells and bubbles with the stellar energy input, one can compare the age and number of supernovae needed to create a bubble (using measurements of size and expansion velocity, and ambient density combined with expansion models) with the ages and number of supernovae expected from clusters that have passed through the same regions. These recent results indicate that multiple supernovae associated with clusters have contributed to bubbles near the Sun, producing a complex environment of time-evolving shells, bubbles, filaments, and secondary/tertiary star forming regions \citep{1974ApJ...189L.105C,2023ASPC..534...43Z}.

However, gaps remain in our ability to fully explain the current structure and ionization state of the local ISM. It is unclear whether the known populations of nearby stellar clusters can account for all observed features, or whether additional sources or mechanisms are required. Moreover, the role of shocks, collisional ionization, and non-equilibrium processes in shaping emission-line signatures is not yet well constrained. These uncertainties complicate efforts to connect observed emission structures to specific past star-forming events. Bridging these gaps is critical for making full use of recent advances in 3D ISM mapping and for building a coherent picture of recent stellar feedback in our Galactic neighborhood.

We have recently carried out three-dimensional Monte Carlo simulations of the radiative transfer of Lyman continuum photons through the local ISM, as characterized by the high-resolution dust map of \citet{edenhofer23}, and demonstrated that a sky projection of the three-dimensional distribution of model H$\alpha$ emissivity is in good agreement with the observations \citep{mccallum25}.  These simulations allow us to reliably determine the temperature and ionization state of the gas throughout the local interstellar medium. In this paper, we present two additional advances that allow us to make a tighter connection between the observed optical emission lines and the evolutionary state of the numerous bubbles and shells seen in the local neighborhood. First, we update our earlier work to include predictions for the emission from a set of common optical and infrared emission lines, e.g. transitions of [N{\sc ii}], [S{\sc ii}], [O{\sc iii}],  [Ne{\sc ii}], and [Ne{\sc iii}]. Secondly, we present a new suite of time-dependent radiation-hydrodynamics simulations using the framework described in \citet{mccallum24a, mccallum24b} to explore the role of shocks, collisional ionization, and non-equilibrium evolution in shaping the optical emission line sky.  We summarize the inputs for our simulations in \S \ref{method_section}, present some of our principal results in \S \ref{results_section} and \S \ref{section-hydroresults}, discuss the implications of this work and comparison with prior results in \S \ref{discussion_section}, and provide a summary of results in \S \ref{conclusion_section}. 

\begin{figure*}
    \centering
    \begin{overpic}[width=\textwidth]{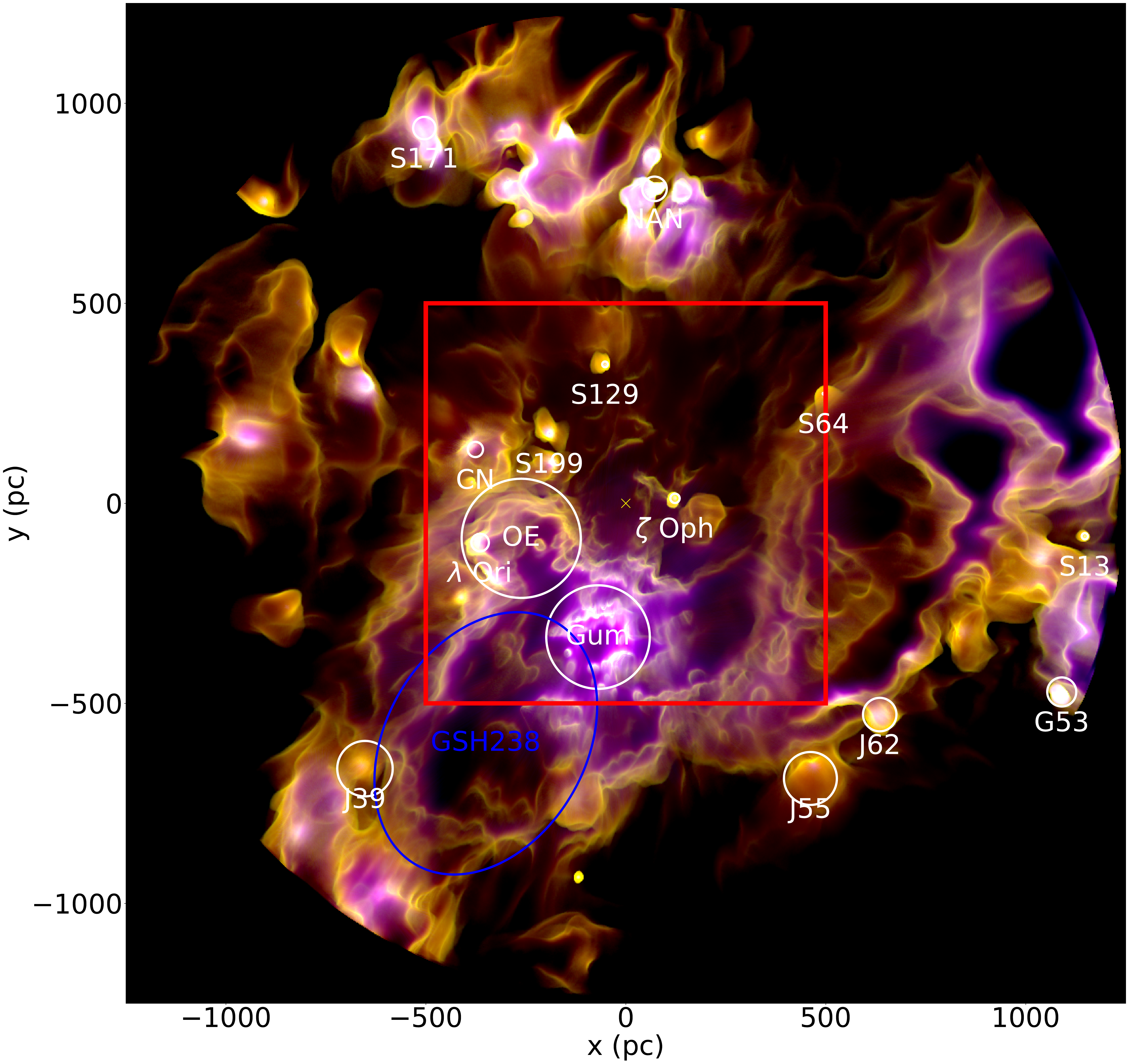}
        \put(82,7){%
            \includegraphics[width=0.17\textwidth]{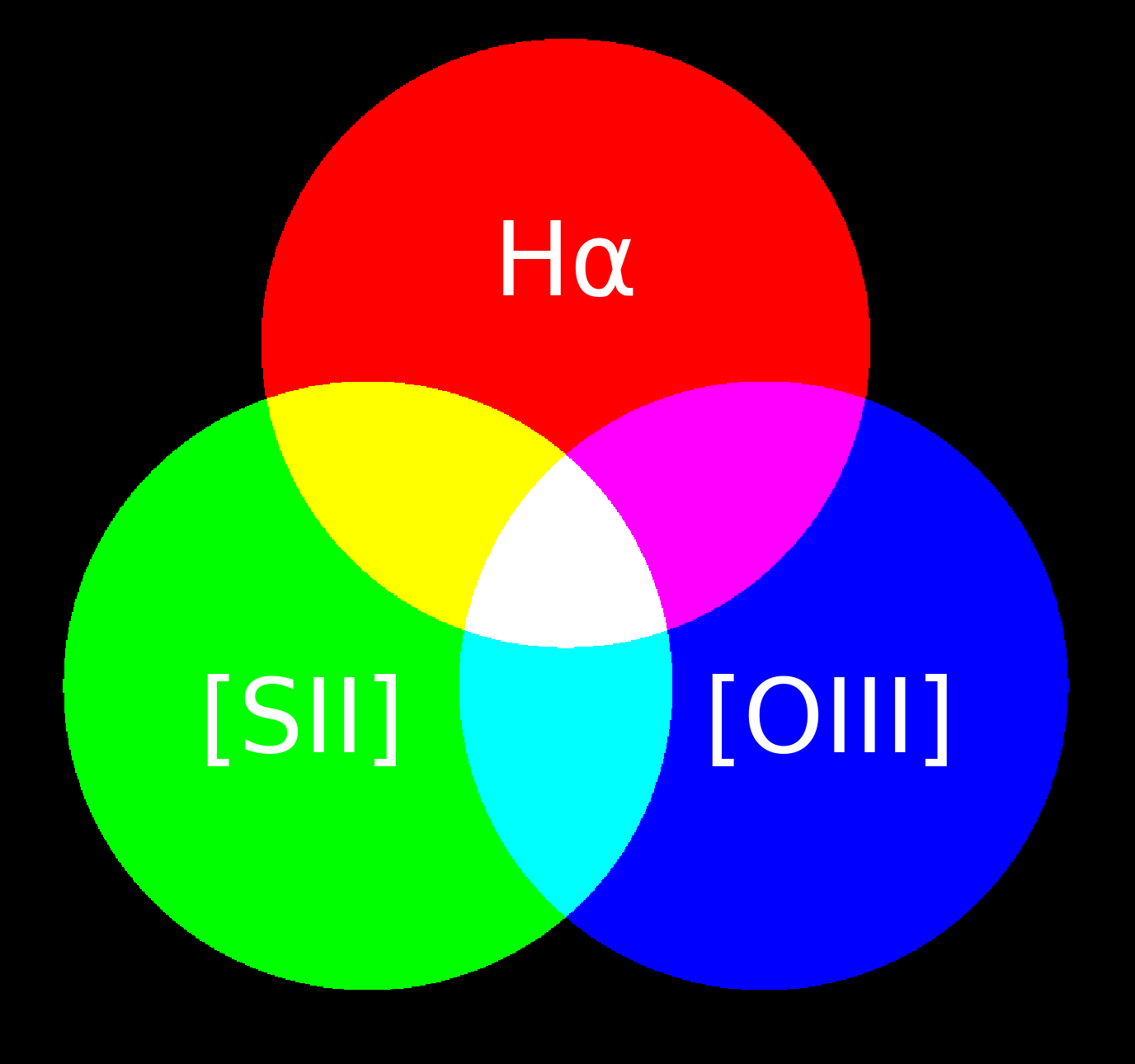}%
        }
    \end{overpic}
    \caption{Face-on view of the \emph{observational} model using the same color scheme as the previous image. A key for the colour scheme is displayed in the bottom right corner, with H$\alpha$ in red, [S{\sc ii}] 6716{\AA} in green, [O{\sc iii}] 5007{\AA} in blue. The same H{\sc ii} regions are annotated in both the sky-view of the previous figure and this face-on view. The low-density dust void associated with Galactic supershell GSH238+00+09 is also shown. The region indicated with the red box will be compared with the hydrodynamical simulations described in section~\ref{section-hydroresults}. }
    \label{fig:faceondust}
\end{figure*}

\section{Methods} \label{method_section}

This paper presents and compares three different models of the local ISM. The first model is a static photoionization simulation of the local density field as mapped by \citet{edenhofer23}, using Lyman continuum (LyC) photons from known O stars as the source of ionization and heating. This is an update of the work presented in \citet{mccallum25}. This will be called the \emph{observational} model, since it is based on the observational constraints of the Milky Way dust maps.  

The second and third models, which we call the \emph{photoionization-only} and \emph{time dependent} models, are the results of a radiation-hydrodynamical simulation that includes star formation, supernova feedback, non-equilibrium ionization of hydrogen, helium, and metal ions, photoionization from O stars, and non-equilibrium heating and cooling as derived from the non-equilibrium metal ionization. These simulations are described further in \citet{mccallum24a,mccallum24b} with some updates noted below. The supernova feedback is implemented using the methods outlined in \citet{gatto15}, whereby an injection radius is calculated as the radius which contains 1000~$\rm M_{\odot}$ of mass. If this radius is smaller than the analytically derived post Sedov-Taylor radius, then momentum is injected, otherwise thermal energy is injected. Photoionization feedback is implemented via the calling of a Monte-Carlo radiative transfer code, which traces Lyman continuum photons through the density grid and adds both the ionizing and heating effect of these photons to the hydro step.

We note that we do not include the effects of rotational shearing, magnetic fields, cosmic rays, and diffuse ionization from the hot gas itself, but these models contain all the physics included in {\it our} hydrodynamic/radiative transfer simulations. Comparable simulations with physical effects beyond those included here are the SILCC project \citep{walch15,rathjen21,rathjen23} and TIGRESS-NCR \citep{kim23}.

For the \emph{photoionization-only} model, we use the density structure of the radiation-hydrodynamics simulation and the location of ionizing sources to re-calculate the thermal and ionization state of the simulation volume.  Therefore, the \emph{photoionization-only} model is identical in construction to the \emph{observational} model, but with the density structure coming from our radiation-hydrodynamical simulations rather than an observational reconstruction.  

For the \emph{time dependent} model, we use the same radiation-hydrodynamical simulation, but preserve the thermal and ionzation structure from the non-equilibrium evolution.  This allows us to examine what kinds of time-dependent structures we might be missing in our \emph{observational} model, since we do not have a full understanding of the time-evolution of the local ISM. 

The radiation transfer for all three types of models were carried out using the \texttt{CMacIonize} code \citep{cmacionize,cmi2} based on the earlier photoionization code from \citet{wood04}. Details are provided below. 

\begin{figure*}
    \centering
    \includegraphics[width=1.0\textwidth]{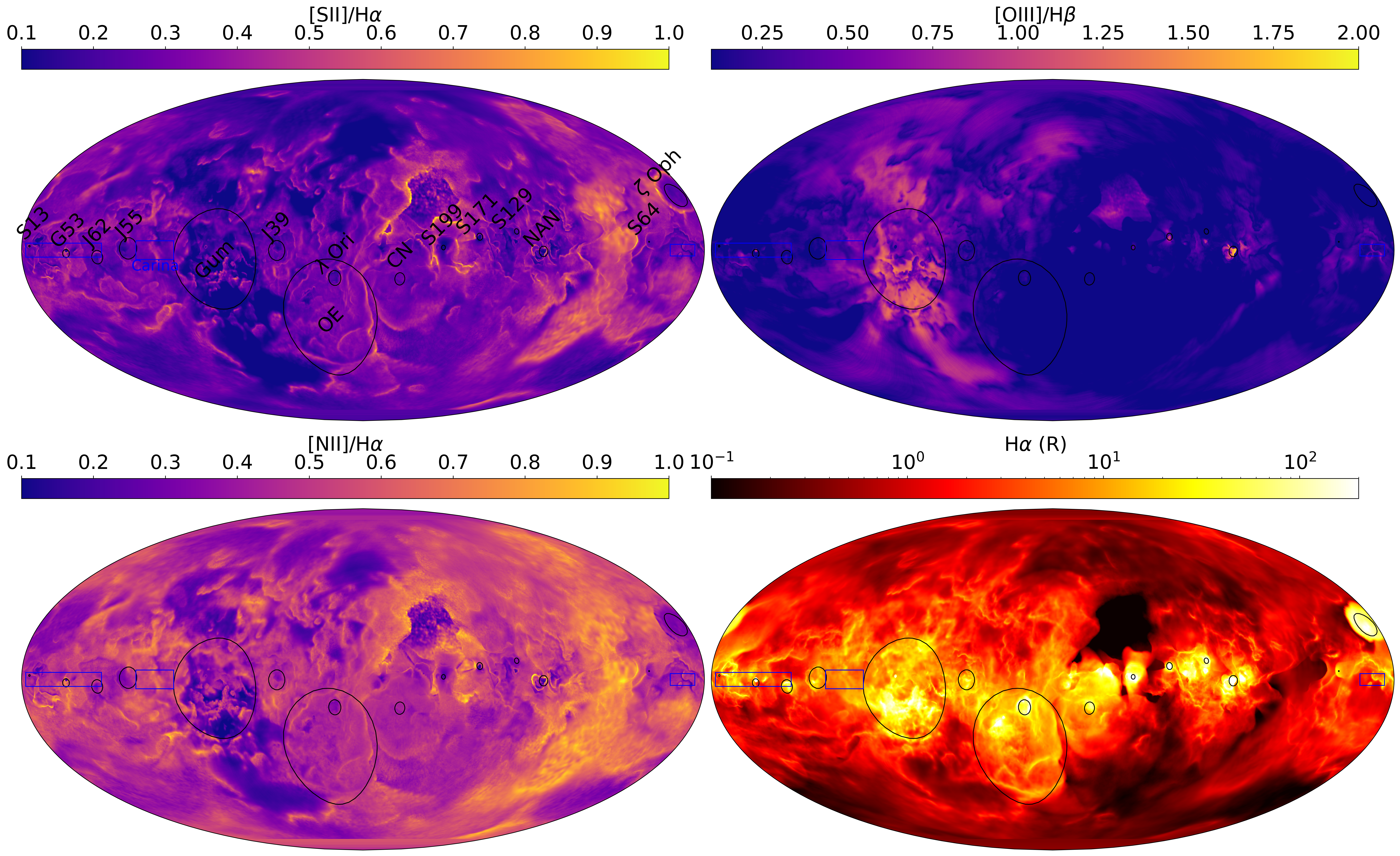}
    \caption{Maps of predicted line ratios projected on the sky with the same annotations and projection as in figure 1. Note that [O{\sc iii}]/H$\beta$ is mostly only seen around the hottest sources, in this case the Bajamar star in the North American Nebula and the very hot stars in the Gum nebula (including the Wolf-Rayet star in the $\gamma^{2}$ Velorum system). [S{\sc ii}]/H$\alpha$ is seen mostly in interface regions between neutral and ionized zones (partially ionized gas), whereas [N{\sc ii}]/H$\alpha$ is seen more to have a more broad and diffuse structure due to the similar ionization potentials on nitrogen and hydrogen. Regions of bright [O{\sc iii}]/H$\beta$ are seen as a deficit of [S{\sc ii}]/H$\alpha$ and [N{\sc ii}]/H$\alpha$. This is due to an excess of high energy photons further ionizing the $\rm N^{+}$ to $\rm N^{++}$, and $\rm S^{+}$ to $\rm S^{++}$.}
    \label{skyratios}
\end{figure*}

\subsection{\emph{Observational} model}
\label{observationalmodel}
As in \citet{mccallum25}, the density grid was derived from the 3D differential dust extinction map of \citet{edenhofer23}. We interpolate the dust map onto a Cartesian cubic grid of $1024\times1024\times1024$ cells, out to a distance of 1.25~kpc in each direction. This gives a simulation box of length 2.5~kpc in each axis. The grid of differential extinction was converted to hydrogen number density using the method described in \citet{mccallum25} (as derived from \citet{zucker21,oniell24}). 

For the distribution of ionizing sources, we use 87 known O stars and one Wolf-Rayet star within 1.25~kpc of the Sun as used in \citet{mccallum25}. Distances were obtained from Gaia parallaxes \citep{gaia} and earlier catalogs where high quality Gaia measurements were unavailable \citep{fabricius02,vanleeuwen07,zacharias12}. 
   
The ionizing luminosities and effective temperatures of each source were interpolated from tables 4-6 from \citet{martins05}, and the appropriate spectrum for each ionizing source taken from the WMBasic library \citep{pauldrach01} as a function of each source's effective temperature. The ionizing luminosity of the WR star was from \citet{crowther07}. 

Calculation of the stellar photoionization is performed using the \texttt{CMacIonize} code in task-based mode,  using the methods described in \citet{vandenbroucke18}. The ionization state of the gas is solved iteratively through a series of Monte Carlo photon shooting steps. For each iteration, the equilibrium temperature structure is calculated with heating from photoionization and cooling from collisionally excited emission lines, thermal bremsstrahlung and losses due to the recombination of ionized hydrogen and helium. Further details on the heating/cooling calculations are shown in appendix~\ref{appendix1}. We solve for the equilibrium multi-level ionization states of oxygen, nitrogen, neon, sulphur and carbon using the mean intensities derived from the final iteration of photon shooting and an electron density which only includes contributions from ionized hydrogen and helium.  The  metal ionization levels also include the effect of collisional ionization and charge transfer to and from hydrogen and helium. The atomic data used were described in \citet{mccallum24b}. We track photons up to energies of 54 eV, providing the 3D ionization structure of all photoionized species up to this ionization potential, as well as a 3D grid of temperature. These are used to calculate the 3D emissivity of collisionally excited forbidden lines as described in \citet{wood04}. 

With this method, the required inputs to run a full simulation are the 3D density structure, the metal abundances, and the distribution of ionizing sources including their ionizing luminosity and spectra. For all simulations presented in this paper we use fixed hydrogen relative abundances of the following; He: 0.1, C: $1.4\times10^{-4}$, N: $7.5\times10^{-5}$, O: $3.19\times10^{-4}$, Ne: $1.17\times10^{-4}$, S: $1.3\times10^{-5}$ \citep{mathis00,wood04b}. We do not account for depletion of metals into dust grains, but we note that the much of the cooling in the warm ionized gas comes from the O{\sc ii} ion, which is not readily precipitated into a solid state.

In addition to the high resolution, 2.5~kpc box run, we run an identical simulation for a smaller cubic box of $\rm 1~kpc \times 1~kpc \times 1~kpc$ comprising $256\times256\times256$ grid cells, still centered on the Sun. This simulation is to allow for direct comparison with the two versions of our radiation-hydrodynamical models which were calculated for the same volume.

\subsection{\emph{Time Dependent} and \emph{Photoionization only} models}

Both the \emph{photoionization} and \emph{time dependent} simulations are the results of time-dependent hydrodynamical simulations coupled with the same radiative transfer methods described above. A full description of the simulation can be found in \citet{mccallum24a, mccallum24b}. These simulations first require a hydrogen-only time-dependent simulation run to 300~Myr, at which point we introduce the full time-dependent metal ionization calculation. In \citet{mccallum24a}, 300~Myr was deemed to be long enough to achieve a quasi-static state. The simulation is then run for an additional 25~Myr to ensure the ionization state is independent of the initial state of the metals when introduced to the simulation. \citet{mccallum24b} found that in general 10~Myr was enough to `forget' the initial conditions of the metal ionization state, however, these simulations focus on the midplane gas and hence involve higher mean densities. This means that our initial ionization states are likely `forgotten' much faster. Simulations longer than 25~Myr could be carried out, but going to longer timescales could introduce non-physical inflow/outflow conditions with the vertical extent of the simulation box having been truncated at $\pm 500~\rm pc$.

Since the focus of our work is on the region covered by the 3D dust maps, we truncate the simulation box vertically and refine the grid when we transition from the hydrogen-only simulations to the full time-dependent metal ionization simulations. The initial hydrogen-only simulations are run with the same resolution and box dimensions as in \citet{mccallum24a}; after truncation and refinement the simulation volume is a cube of $1~\rm{kpc}$ in each dimension with a resolution of 3.9~pc. Although this volume is not enough to follow the inflow and outflows described in our previous work, it is sufficient to produce a grid of ionization states for comparison with the \emph{observational} model described above. We estimate using previous tall-box simulations that our restricted volume still contains 96\% of the H$\alpha$ emissivity of the full $|z| < 3~\rm kpc$ simulation volume.

 In addition, as compared to \citet{mccallum24a}, we explore a wider range of star formation rates.  In our original work, the star formation rate within the tall box simulations was calculated self-consistently as a function of the gas mass within 250~pc of the midplane (consistent with the Kennicutt-Schmidt relation). This  relies on an initial star formation rate value to be chosen, representing the normalisation of the Kennicutt-Schmidt relation. In our previous works, an initial value of 1200~$\rm M_{\odot}~\rm{Myr}^{-1}~\rm{kpc}^{-2}$ was used for fiducial simulations. This was motivated by our efforts to match the observed one kiloparsec scale height of diffuse ionized gas without destroying the neutral disc.

Here our focus is on finding the level of star formation needed to match the morphology of the warm ionized medium as characterized in our \emph{observational} model. So we explore a range of star formation rates by running four different simulations with initial SFRs of 185, 370, 740 and 1480~$\rm M_{\odot}~\rm{Myr}^{-1}~\rm{kpc}^{-2}$.

There are three other departures from the procedure followed in \citet{mccallum24a,mccallum24b}.  First, the hydrogen-only simulations have been updated to use a fourth order Runge–Kutta ODE solver rather than the ionization state `limiter' method described in \citet{mccallum24a}. Second, we include a wider range of O star spectra, unlike the single 40kK spectrum used for all O stars in \citet{mccallum24b}. And third, we modify the \emph{molecular cloud ionizing photon escape fraction} used in this previous work. This is a corrective multiplicative factor for the ionizing luminosity used for each star. In \citet{mccallum24a} this factor was set to 0.1 to account for photon losses to unresolved maxima in the density structure. But in order to maintain consistency with the \emph{observational} model derived from the 3D dust maps, we set this factor to one for all simulations. 

{\it Time dependent} model:  In the \emph{time dependent} model, the temperatures, densities, and non-equilibrium ionization balance obtained from the simulations described above are used to calculate the 3D emissivities of all of the optical emission lines of interest. In particular, this includes the [O{\sc iii}] and [Ne{\sc iii}] emission lines which are strongest in regions of high-excitation (near hot stars) or in regions that are collisionally ionized.  

{\it Photoionization-only} model: In the ``photoionization-only'' models, the final density structure of the simulation above is combined with the locations and stellar masses of the ionizing sources at the end of the 25 Myr simulation in order  to calculate 3D grids of emissivity using the same photoionization code described above.  In this case, we replace the temperatures and ionization balances from the time-dependent code with new values that include only the effects of stellar photoionization equilibrium. The spectrum of each source is determined by interpolating the main sequence table from \citet{martins05} to identify an effective temperature from a stellar mass, at which point a WMBasic \citep{pauldrach01} spectrum can be assigned to each source. Since our photoionization code only includes photons of energy less than 54 eV, ions with a higher ionization potential are absent in this model.

\begin{figure*}
    \centering
    \includegraphics[width=0.8\textwidth]{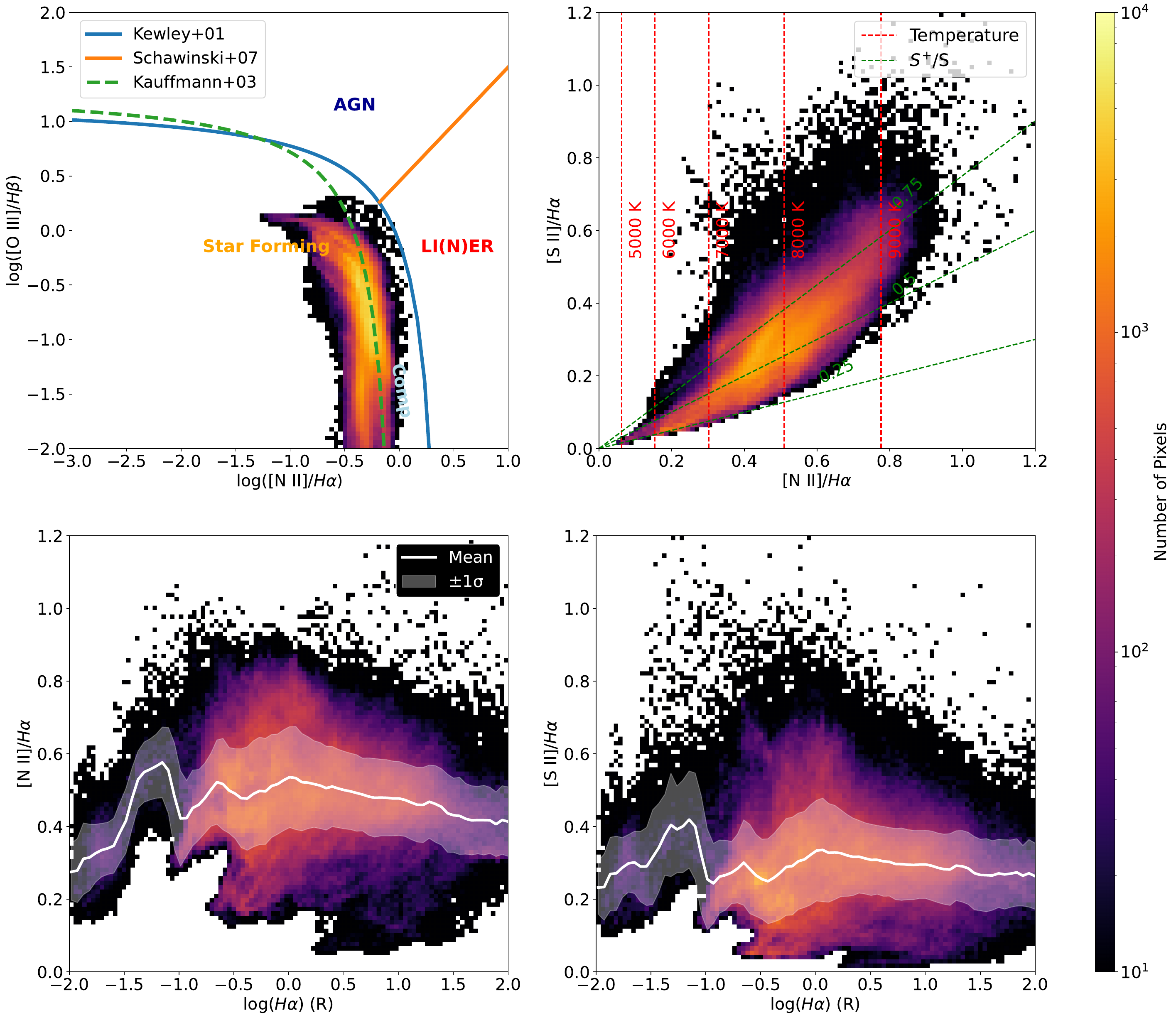}
    \caption{Diagnostics from the simulated sky as generated from the \emph{observational} simulation. HEALPix skies were generated, and the HEALPix were then binned in the 2D plane of each emission line diagram, with brighter regions denoting more HEALPix producing that set of emission lines.
    Top left panel is the [N{\sc ii}]/H$\alpha$ versus [O{\sc iii}]/H$\beta$ BPT diagram. Top right is the diagnostic diagram of [N{\sc ii}]/H$\alpha$ versus [S{\sc ii}]/H$\alpha$. The bottom left panel shows the variation of [N{\sc ii}]/H$\alpha$ with H$\alpha$ intensity and bottom right shows [S{\sc ii}]/H$\alpha$ versus H$\alpha$ intensity. The bottom panels also show the mean line ratio and dispersion in each bin of H$\alpha$ surface brightness. Note in the BPT diagram, all pixels fall below the \citet{kewley01} `maximum starburst line' as expected in a simulation of \emph{photoionization only}. Note also the mean value of [S{\sc ii}]/H$\alpha$ in all bins never exceeds 0.4, cited by \citet{caldwell25} as an oft-used lower limit for the detection of supernova remnants.}
    \label{fig:skydiag}
\end{figure*}

\section{Results of the \emph{Observational} Model} \label{results_section}

In this section, we will first present a sky view and face-on view of the optical emission line results for our \emph{observational} model (a multi-wavelength update of \citet{mccallum25}) followed by an analysis of emission line diagnostics. The following section provides the results of our radiation-hydrodynamical simulations, both the \emph{time dependent} model and the \emph{photoionization-only} model. 

\subsection{The Multi-Wavelength Sky in 3D}

In \citet{mccallum25}, we presented the H$\alpha$ sky as predicted from this simulation, and noted the good agreement with the observed H$\alpha$ sky as measured by the Wisconsin H$\alpha$ Mapper \citep{2003ApJS..149..405H}. Figure~\ref{linepanels} updates this result with a three-channel RGB image, where red is H$\alpha$, green is [S{\sc ii}] 6716{\AA} and blue is [O{\sc iii}] 5007{\AA}. This RGB color scheme is utilized throughout this paper. Figure~\ref{linepanels} includes the effects of dust scattering and attenuation, while all other figures display collapsed grids of emissivity without dust effects.  

The sky in [S{\sc ii}] 6716{\AA} and [N{\sc ii}] 6584{\AA} bears a morphological resemblance to the sky in H$\alpha$, whereas [O{\sc iii}] 5007{\AA} (and [Ne{\sc iii}] 15~$\mu m$, which is not shown) are only spatially coincident with particular regions. These regions of O{\sc iii} and Ne{\sc iii} are confined to the immediate surroundings of only the most massive/hottest sources with the hardest spectra. The most notable extended region of these higher energy ions is around the Gum nebula, which contains two very luminous sources, including the $\gamma^{2}$ Velorum system with a luminous and spectrally hard Wolf-Rayet source.

In contrast, the $\zeta$ Oph H{\sc ii} region, which appears brightly in H$\alpha$, N{\sc ii} and S{\sc ii}, is hardly seen in O{\sc iii} and not at all in Ne{\sc iii} due to the relative softness of the spectrum from the O9.2 star at its core.

The Orion-Eridanus superbubble also seems to be, on average, lower excitation than the Gum Nebula, but with with some small pockets of stronger  O{\sc iii} emission.  One of these pockets is the $\lambda$ Ori (O8III) H{\sc ii} region, which we have shown to be a blister H{\sc ii} region with the ionizing source existing in a dimple-like structure at the back of the Orion-Eridanus superbubble.

Figure~\ref{fig:faceondust} shows a face-on view of our local region of the Milky Way, as would be viewed by an external observer. This image is a three-channel colour image using the same scheme as in the previous figure. In this image we see bubbles and voids, as well as apparently distinct, dense and bright H{\sc ii} regions. When an ionizing source exists in a particularly dense region of the dust map, a classical ionization bounded H{\sc ii} region is formed (see the $\zeta$ Oph H{\sc ii} region in \citet{mccallum25}). When bright ionizing sources exist in the larger, low density regions of the grid, the photons are caught by the inner surface of the void, where gas has piled up into shell like structures. Many of the voids are connected via holes and channels, creating a complex system of ionized surfaces. Higher brightnesses of [O{\sc iii}] emission are seen on the inner layer of these surfaces, giving way to lower ionization energy ions as we move deeper into the gas, shielded from Lyman continuum photons.

\subsection{Optical Emission Line Diagnostic Diagrams}

Figure~\ref{skyratios} shows sky maps of the line ratios of [N{\sc ii}]/H$\alpha$, [S{\sc ii}]/H$\alpha$ and [O{\sc iii}]/H$\beta$. [S{\sc ii}] is mostly seen in the partially ionized interface regions; areas of the sky which show high values of [S{\sc ii}]/H$\alpha$ have a filamentary structure. This differs from the [N{\sc ii}]/H$\alpha$ map which is more diffuse, due to the similarities of the ionization potentials of nitrogen and hydrogen. [O{\sc iii}]/H$\beta$ is elevated in only a few regions of the sky, most notably within the Gum nebula and the North American nebula, which contain the two most luminous and spectrally hard O stars in the simulation: $\zeta$ Puppis and the Bajamar star, respectively. Regions of bright [O{\sc iii}]/H$\beta$ are seen as deficits of [N{\sc ii}]/H$\alpha$ and [S{\sc ii}]/H$\alpha$. This is due to very energetic photons that can ionize $\rm N^{+}$ to $\rm N^{++}$ and $\rm S^{+}$ to $\rm S^{++}$.

Figure~\ref{fig:skydiag} shows diagnostic optical emission line diagrams from these multi-wavelength calculations, including [N{\sc ii}]/H$\alpha$ versus [O{\sc iii}]/H$\beta$ (BPT diagram as decribed by \citet{baldwin81}), [N{\sc ii}]/H$\alpha$ versus [S{\sc ii}]/H$\alpha$, and the variation of the [N{\sc ii}]/H$\alpha$ and [S{\sc ii}]/H$\alpha$ line ratios with H$\alpha$ intensity. These diagrams were created by summing the simulated line intensities along HEALPix pixels.

In the [N{\sc ii}]/H$\alpha$ versus [O{\sc iii}]/H$\beta$ BPT diagram, all pixels fall below the \citet{kewley01} `maximum starburst line,' in line with expectations in a simulation containing only photoionization from young stars. The [N{\sc ii}]/H$\alpha$ versus [S{\sc ii}]/H$\alpha$ shows the bulk of the pixels occupy a region surrounding [N{\sc ii}]/H$\alpha$ of 0.45 and [S{\sc ii}]/H$\alpha$ of 0.25. Using the same analytical model as \citet{madsen06}, which characterizes these ratios with an electron temperature and sulfur ionization fraction, these simulated data increase roughly linearly with an inferred S$^+$/S of 0.5 and 0.75. With the bulk of [N{\sc ii}]/H$\alpha$ being around 0.45, this implies mean DIG temperatures from use of the diagnostic diagram of 7000-8000~K. The mean temperature of all gas in the simulation with a neutral fraction below 0.5 is 7400~K, supporting the validity of this diagnostic diagram as a temperature measure in purely photoionized gas.

The diagrams of [N{\sc ii}]/H$\alpha$ and [S{\sc ii}]/H$\alpha$ versus H$\alpha$ intensity appear morphologically similar to each other, with a roughly constant value of [N{\sc ii}]/H$\alpha$ and [S{\sc ii}]/H$\alpha$ as a function of H$\alpha$ intensity between 1 and 100~R of H$\alpha$ intensity, with a slight tendency for lower line ratios at higher H$\alpha$ surface brightness. Below an H$\alpha$ surface brightness of 0.1~R, this trend is disrupted, likely due to the poor HEALPix statistics (with so few pixels showing surface brightnesses this low). This range of H$\alpha$ surface brightness is likely below the observational limits of large scale surveys such as the Wisconsin H$\alpha$ Mapper.

\begin{figure*}
    \centering
    \includegraphics[width=1.0\textwidth]{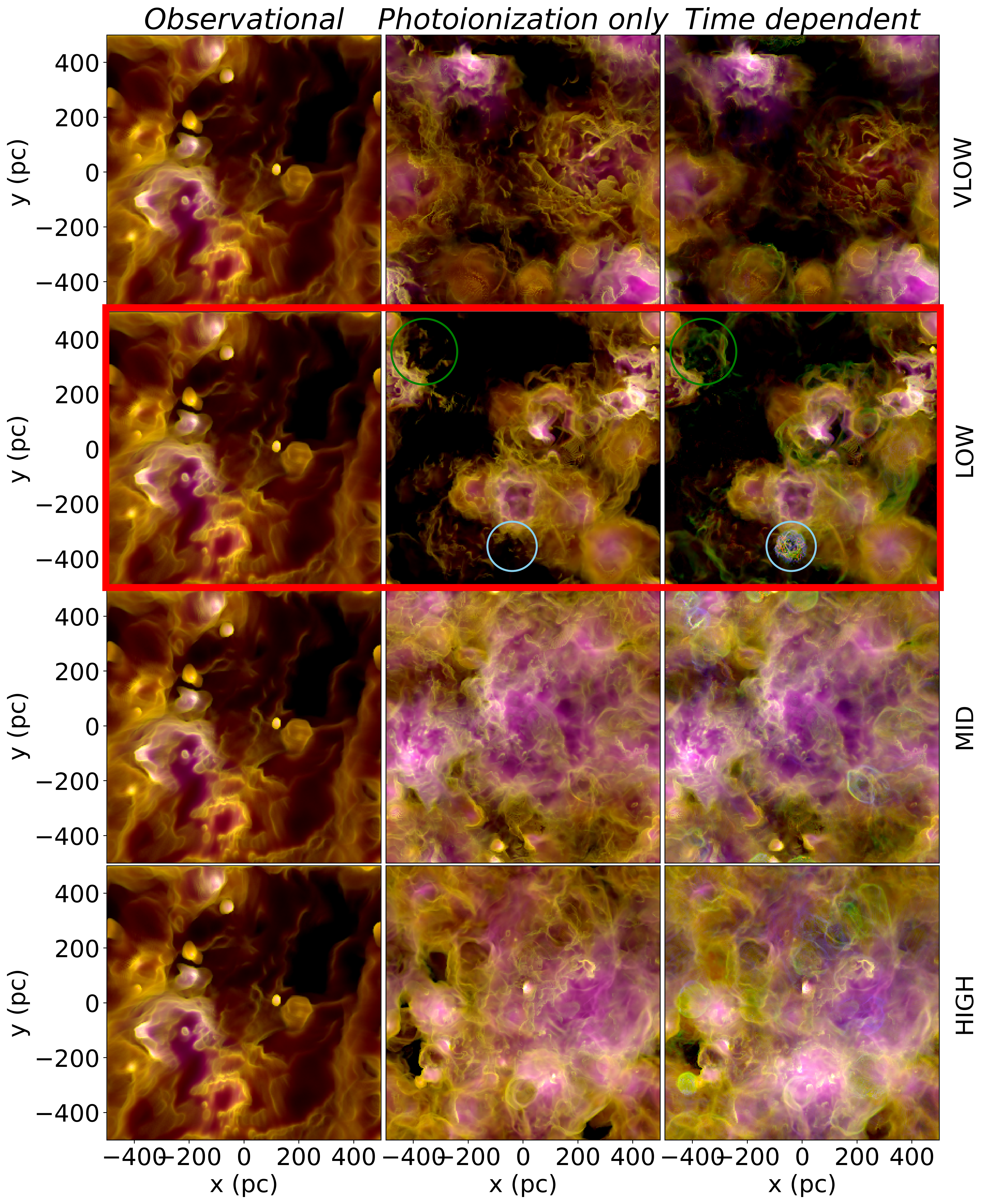}
    \caption{Face-on views of each simulation for four values of the star formation rate (185, 370, 740 and 1480 $\rm M_{\odot} \rm Myr^{-1}kpc^{-2}$) with the star formation rate increasing with each subsequent row.  The left column shows the static dust map simulation (repeated each time), the middle column the post-processed \emph{photoionization-only} model, and the right column the \emph{time dependent} model. The LOW star formation rate provides the best match between the \emph{observational} and \emph{photoionization-only } models. Note that the \emph{time dependent} model contains blue and green bubbles (marked with circles) that are characterized by bright [O{\sc iii}] and [S{\sc ii}] emission, showing the location of young supernova remnants. These objects are not seen in the \emph{photoionization-only} models. See text for further description and comparison of these models.}
    \label{rgbs}
\end{figure*}

\section{Results of the Radiation-Hydrodynamical Simulations} \label{section-hydroresults}

\subsection{Matching the \emph{photoionization-only} models with the \emph{observational} model}

Using the methods described in \S \ref{method_section}, we have calculated the {\it ab initio} evolution of a kiloparsec-sized cube of the ISM for four values of the star formation rate: 185 (VERY LOW), 370 (LOW), 740 (MID), and 1480 (HIGH) $\rm M_{\odot}~\rm{Myr}^{-1}~\rm{kpc}^{-2}$. 
Figure~\ref{rgbs} shows the face-on view for a single snapshot in time of ISM morphology using the same color-table for the H$\alpha$, [S{\sc ii}] 6716{\AA}, and [O{\sc iii}] 5007{\AA} emission as in \S \ref{results_section}. The different levels of star formation are shown in each row. The centre column shows a visualization of these simulations using the \emph{photoionization-only} model, and the right column shows the \emph{time dependent} model.  For convenience, the \emph{observational} model  for the kiloparsec around the Sun is repeated in each of the four rows. Properties of each of the models for the selected snapshot, including the number and ionizing luminosity of O stars and total intensity of selected optical and infrared emission lines are given in Table~\ref{simcompare}. The line intensities are given for the \emph{photoionization} models only, because we are using these models to compare to the \emph{observational} results.

The \emph{photoionization-only} simulation snapshots bear the best qualitative resemblance to the dust map simulation at star formation rates of 185 and 370~$\rm M_{\odot}~\rm{Myr}^{-1}~\rm{kpc}^{-2}$. At higher star formation rates, the supernovae disrupt the density structure beyond what is seen in the dust map derived results. In addition, the total ionizing luminosity in the simulation volume almost completely ionizes the midplane, and generates an extended quantity of [O{\sc iii}] emission, well beyond what is predicted by the \emph{observational} model of the local region.

\begin{table*}
    \centering
\resizebox{\textwidth}{!}{%
\begin{tabular}{llrrrrrrr}
\hline
 Name    & $\Sigma_{\rm SFR}$ ($\rm M_{\odot}~Myr^{-1}~kpc^{-2}$)    &   $QH_{0}/(10^{49}~\rm s^{-1})$ &   $N_{\rm sources}$ &   $I_{\rm H\alpha}$ &   $I_{[\rm OIII}]$ &   $I_{[\rm SII]}$ &   $I_{[\rm NII]}$ &   $I_{[\rm NeIII]}$ \\
\hline
 \emph{Observational} simulation & N/A &  8.913 &          14 &     1.06e+38 &   3.8e+36  &  3.12e+37 &  4.42e+37 &    2.21e+36 \\
\hline
 VLOW    & 185  &  7.353 &          10 &     7.87e+37 &   1.58e+37 &  2.36e+37 &  3.35e+37 &    9.66e+36 \\
 LOW     & 370   &  7.961 &          14 &     9.32e+37 &   9.02e+36 &  3.47e+37 &  4.42e+37 &    5.47e+36 \\
 MID     & 740   & 18.359 &          32 &     2.06e+38 &   3.83e+37 &  5.91e+37 &  9.17e+37 &    2.24e+37 \\
 HIGH    & 1480   & 20.645 &          43 &     2.48e+38 &   2.8e+37  &  7.04e+37 &  1.17e+38 &    1.64e+37 \\
\hline
\end{tabular}
    }
    \caption{Properties of the \emph{photoionization only} simulations for various star formation rates in comparison to the \emph{observational} simulation snapshot. Intensities are total intensity of the entire $1~\rm kpc^{3}$ simulation volume in units of $\rm erg ~s^{-1}$.}
    \label{simcompare}
\end{table*}

As can be seen in table~\ref{simcompare}, the LOW SFR simulation best matches the \emph{observational} model in terms of number of ionizing sources, total ionizing luminosity, as well as H$\alpha$, [S{\sc ii}] and [N{\sc ii}] intensity.  However, this model somewhat overproduces the observational reconstruction of  [O{\sc iii}] and [Ne{\sc iii}] intensity by around a factor of 3. Since these higher excitation lines are sensitive to the small number of highest mass stars we believe that the LOW SFR provides the best match to the data; this match could be improved  by artificially steepening the rad-hydro IMF to produce the observed lower levels of [O{\sc iii}] and [Ne{\sc iii}].

\subsection{The effects of shocks and non-equilibrium ionization}

A comparison of the middle and right columns of figure~\ref{rgbs}, which both derive from the same radiation-hydodynamical simulations, illustrate the differences between the \emph{photoionization-only} and \emph{time dependent} multiwavelength view and provide insights to what might be missing in our observational reconstruction of the local ISM.  Two of the principal differences seen in the two views are bubbles of bright  [O{\sc iii}] and [S{\sc ii}], marked with blue and green circles, respectively. These bubbles mark the location of relatively hot post-shock gas not captured in the \emph{photoionization-only} snapshots. The bright [O{\sc iii}] bubbles are the younger supernova remnants, where the gas in the front is in a higher state of ionization (oxygen is doubly ionized due to collisional ionization in the hot bubble front). This [O{\sc iii}]-bright phase is much shorter lived, typical lifetimes of around 0.3 Myr, than the later [S{\sc ii}] bright phase which typically lasts approximately 2~Myr. Regions of bright [O{\sc iii}] are also seen around some of the more massive/spectrally hard sources in the simulation, but these regions also have strong H$\alpha$ so they appear magenta in our images (bright [O{\sc iii}] and H$\alpha$ with lower [S{\sc ii}]). 

Regarding the bright [S{\sc ii}] bubbles, we note that singly ionized sulphur (S$^+$) is the lowest ionization state tracked by the code. Sulphur has a first ionization potential of 10.4 eV, below the Lyman continuum threshold. Therefore, these `green' structures in [S{\sc ii}] emission are not due to the gas being highly ionized, but rather are tracing regions that are closer to `neutral' but still hot following the passage of the supernova shock.  The density fronts that are bright in [S{\sc ii}] have temperatures around $10^{4}~\rm K$. H$\alpha$ is also present in these fronts, but with lower emissivity due to the inverse relationship between temperature and H$\alpha$ brightness ($\rm \epsilon_{H\alpha} \propto T^{-0.938}$ \citep{osterbock06}).  The traditional optical emission line ratio used to identify supernova remnants in other galaxies is [S{\sc ii}]/H$\alpha> 0.4$ \citep{caldwell25} which is what we observe in these simulations. 

\begin{figure}
    \centering
    \includegraphics[width=0.8\linewidth]{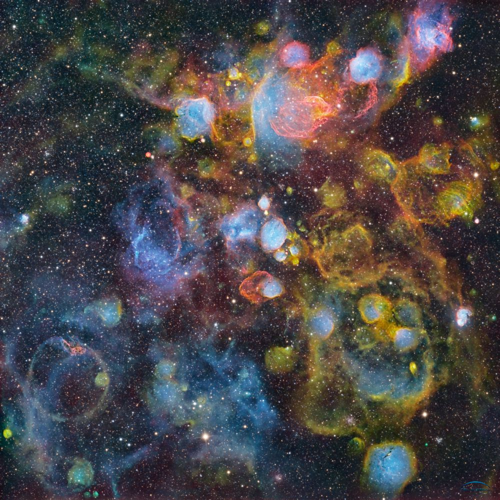}
    \caption{The Small Magellanic Cloud as observed by the collective of amateur astronomers, `Ciel Austral' (\href{https://www.cielaustral.com/}{https://www.cielaustral.com/}). This is a 3-channel false colour image as prepared in figure~\ref{rgbs}, with the same colour scheme applying to the same emission lines, that is red H$\alpha$, green [S{\sc ii}] 6716{\AA}, and blue [O{\sc iii}] 5007{\AA}. The extent of this image in physical space is around 500~pc in each axis, and is thus comparable to one quadrant of any of the face-on panels in figure~\ref{rgbs}.}
    \label{smc}
\end{figure}

One omission in our \emph{time dependent} models is that we do not include the effects of X-ray and UV radiation from post-shock gas. SNRs are known to lose a significant amount of energy to Lyman continuum photons \citep{makarenko23}. While these energy losses are included in our heating/cooling prescription, this emission is also known to contribute to the ionization of the post-shock gas. This is especially true in post-shock regions with magnetic fields that inhibit the compression of the post-shock gas and increase the effectiveness of photoionization. Further development of the \texttt{CMacIonize} code would be required to implement the additional ionization effect of the hot gas. 

Our fiducial resolution of 3.9 pc does not fully resolve radiative shocks and the detailed non-equilibrium structures that may develop within them. However, we note that the broad non-equilibrium emission features observed in our simulations are not isolated to narrow, shock-confined layers but rather arise from more extended supernova-heated gas. While we acknowledge that some finer shock-driven structures are unresolved, the main features we attribute to non-equilibrium effects are spatially extended and not dominated by the smallest resolved scales.

The parsec-resolution face-on view we are able to generate for comparison with Milky Way data are most comparable to galaxies in the Local Group for which we can obtain observations with a similar angular resolution. As a comparison to our simulations, we include a three-colour image of the Small Magellanic Cloud (SMC) in the same colour scheme as figure~\ref{rgbs}. This image was observed by the collective of amateur astronomers, `Ciel Austral'. \footnote{{https://www.cielaustral.com/}} This image spans around 500~pc, and similar structures to the simulations can be identified, including near spherical shells of [O{\sc iii}] emission, bright rings of H$\alpha$, and numerous linear structures emitting strongly in lines of [S{\sc ii}]. Whilst this image of the SMC provides an interesting visual, we caveat the comparison to the local Milky Way with the reality of differing interstellar properties. The SMC has a lower dust content, has one quarter the metallicity and a much higher SFR than the local ISM. These differences will impact the way in which stars are able to feedback into their environment.

\subsection{Diagnostic Diagrams}

\begin{figure}
    \centering
    \includegraphics[width=1.0\linewidth]{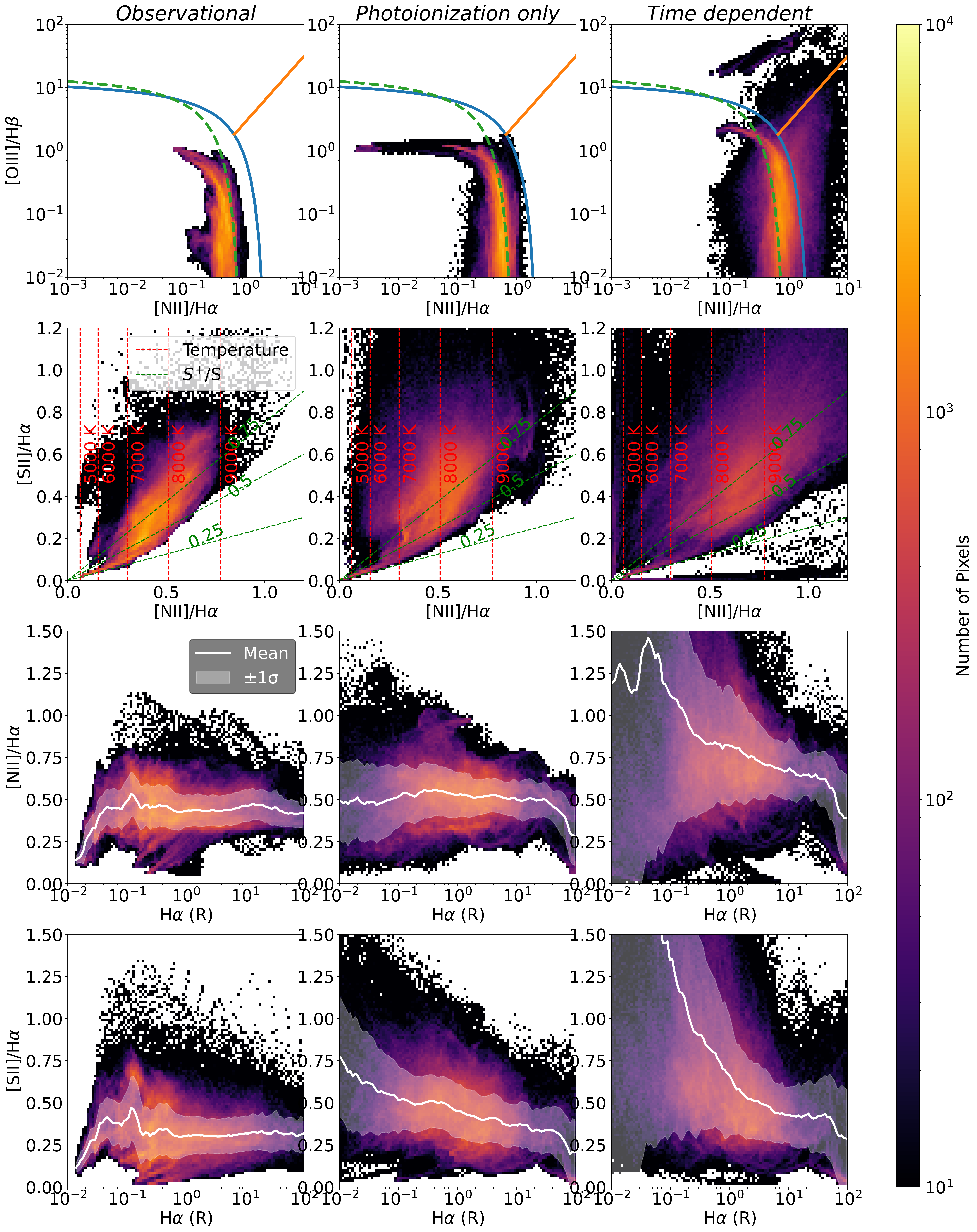}
    \caption{Diagnostic diagrams formed from synthetic skies from both the \emph{observational} simulation, and the \emph{photoionization only} LOW star formation rate run, and the \emph{time dependent} LOW star formation rate run. The left column shows the \emph{observational} simulation, the middle column shows the \emph{photoionization-only} results from the radiation-hydrodynamics simulation, and the right column shows the same snapshot from the \emph{time dependent} simulation.}
    \label{bpts}
\end{figure}

Figure~\ref{bpts} shows the resulting diagnostic diagrams from our LOW star formation rate run alongside the \emph{observational} simulation. These diagrams were produced by creating a synthetic HEALPIX sky from the central voxel of the simulation, and creating a 2D histogram, where each bin is counting the number of HEALPIX pixels. While this pixel-based weighting reflects the frequency of occurrence of line ratios across the sky, it does not account for the relative luminosity contribution of each region. We find that our main conclusions are insensitive to the weighting scheme, though we note that alternative weightings — such as by H$\alpha$ luminosity — may be important for studies focused on the brightest or most observationally dominant regions.

The \emph{photoionization only} and \emph{observational} simulations appear similar in the BPT diagrams (top row), with all of the pixels falling below the Kewley \emph{maximum starburst line}. This is in line with expectation, as these simulations contain only emission from photoionized gas from OB stars, no spectrally hard evolved sources, no AGN, and no supernova feedback. For the \emph{time dependent} simulation which also includes supernova feedback, the bulk of the emission moves up and to the right in the BPT diagram, representing a hardening of [N{\sc ii}]/H$\alpha$ and [O{\sc iii}]/H$\beta$. [N{\sc ii}]/H$\alpha$ is hardened due to the increase in temperature from the additional thermal energy from the SNe, and [O{\sc iii}]/H$\beta$ is hardened due to collisional ionization adding to the high energy species of O{\sc iii}. We also see a large number of cells appearing to the right of the \emph{maximum starburst line}, in the region of the diagram defined as LI(N)ER gas (low ionization nuclear emission region). This was also seen in the face-on BPT diagrams from \citet{mccallum24b}, in which the introduction of supernova feedback and time-dependent ionization generated gas which was emitting in the LI(N)ER section of the BPT diagram. This LI(N)ER categorized gas is also seen in the data of \citet{caldwell25}, identifying many SNRs in M31 in this region of the BPT diagram. This may suggest that the BPT diagram is in practice unable to disentangle the ionization mechanism of photoionization by spectrally hard sources (subdwarf OB stars or other), and collisional ionization through the thermal energy injection from supernovae.

The [N{\sc ii}]/H$\alpha$ versus [S{\sc ii}]/H$\alpha$ diagnostic diagram is shown the second row of figure~\ref{bpts}. The \emph{photoionization-only} snapshot shows both higher values of [N{\sc ii}]/H$\alpha$ and [S{\sc ii}]/H$\alpha$ than the \emph{observational} simulation. This indicates higher electron temperatures (8200~K compared to 7300~K) for sightlines in the hydrodynamical simulations as compared to the observational reconstructions. The \emph{time dependent} snapshot shows a further elevation of both line ratios, and a further increase in inferred temperatures, due to the introduction of heating from supernovae shock waves.  

In the bottom two rows of figure~\ref{bpts} are [N{\sc ii}]/H$\alpha$ and [S{\sc ii}]/H$\alpha$ as a function of H$\alpha$ brightness. Again, we see only subtle differences between the \emph{observational} and \emph{photoionization only} simulations, but on the introduction of supernovae and time-dependent ionization we see both of these plots gain a steep rise in line ratios with decreasing H$\alpha$ surface brightness. This is in line with observations of the Milky Way's DIG as presented in \citet{haffner99} and \citet{madsen06}. In the \emph{photoionization-only} simulation, only a slight rise is seen in [S{\sc ii}]/H$\alpha$ at low H$\alpha$, with [N{\sc ii}]/H$\alpha$ remaining mostly flat. This is in line with the results of \citet{mcclymont24}, who post-processed an isolated galaxy simulation under the assumption of ionization equilibrium. The absence of this rise in the \emph{photoionization-only} simulations might suggest that this could be a feature of the inclusion of SNRs and collisionally ionized gas. 

These models will be compared to multi-wavelength studies of warm ionized medium from WHAM and the Local Volume Mapper when these data become available.

\subsection{Importance of O-star environment}

\begin{figure*}
    \centering
    \includegraphics[width=1.0\linewidth]{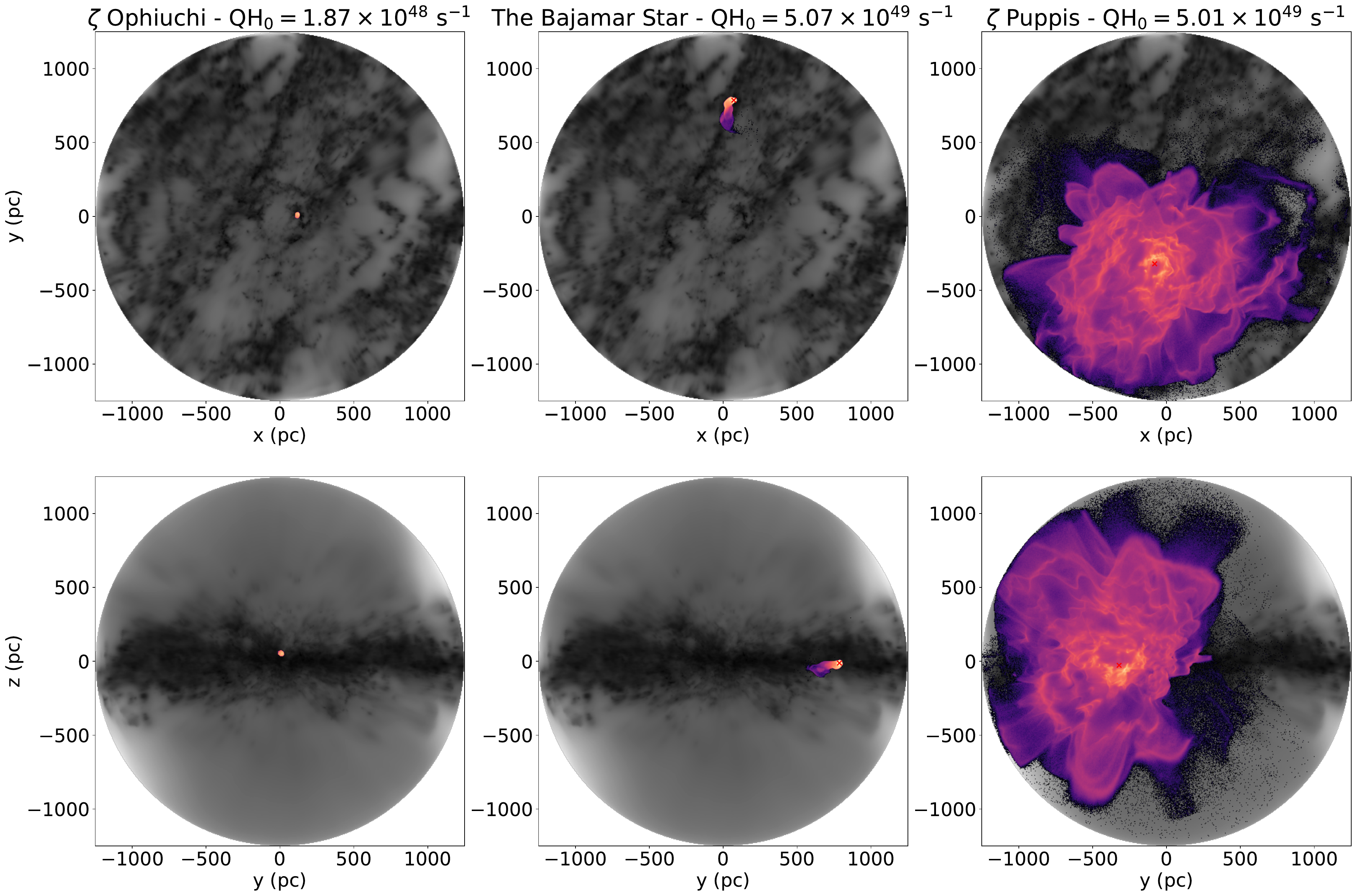}
    \caption{A projection showing the column density plot of where the Lyman continuum photons from individual sources are terminated. Underlying the coloured volume of each zone of influence is the column density of the total mass in the simulation. The top row shows face-on projections of the $x-y$ plane, and the bottom row shows edge-on projections of the $y-z$ plane. The left column shows the region of influence of $\zeta$ Ophiuchi, the middle column shows the Bajamar star, and the right column shows $\zeta$ Puppis in the Gum nebula. A red cross shows the location of each source, but has been omitted on the $\zeta$ Ophiuchi column to avoid obscuring the region. Despite having near equivalent ionizing luminosities, the difference in the volume influence of $\zeta$ Puppis and the Bajamar star is a factor of 827.}
    \label{boundedness}
\end{figure*}

One of the crucial aspects in any individual O-star's ability to ionize a large volume of diffuse ionized gas is the density structure in which it resides. To demonstrate this, we focus on 3 different O-stars in the \emph{observational} simulation; $\zeta$ Ophiuchi, the Bajamar star and $\zeta$ Puppis. To investigate the spatial extent of the influence of each source, we implement a photon packet counter to the MCRT simulation which counts the termination of each photon packet's journey, binned by the physical location of each termination event. These three sources were chosen as they represent a broad spectrum of ionization `boundedness.' $\zeta$ Ophiuchi forms a nearly spherical, ionization bounded H{\sc ii} region from which no LyC photons escape. The Bajamar star forms a similar structure, but with LyC photons `breaking out' in one direction. $\zeta$ Puppis (in the Gum nebula) is situated in an environment which is `leaky' enough that LyC photons are able to travel kiloparsecs of distance and ionize a volume which is 827 times that of the Bajamar star, despite having near equivalent ionizing luminosities. Figure~\ref{boundedness} shows the ionized volume of each of these stars, both from a face-on and edge-on view.

\begin{figure*}
   \centering
   \includegraphics[width=1.0\textwidth]{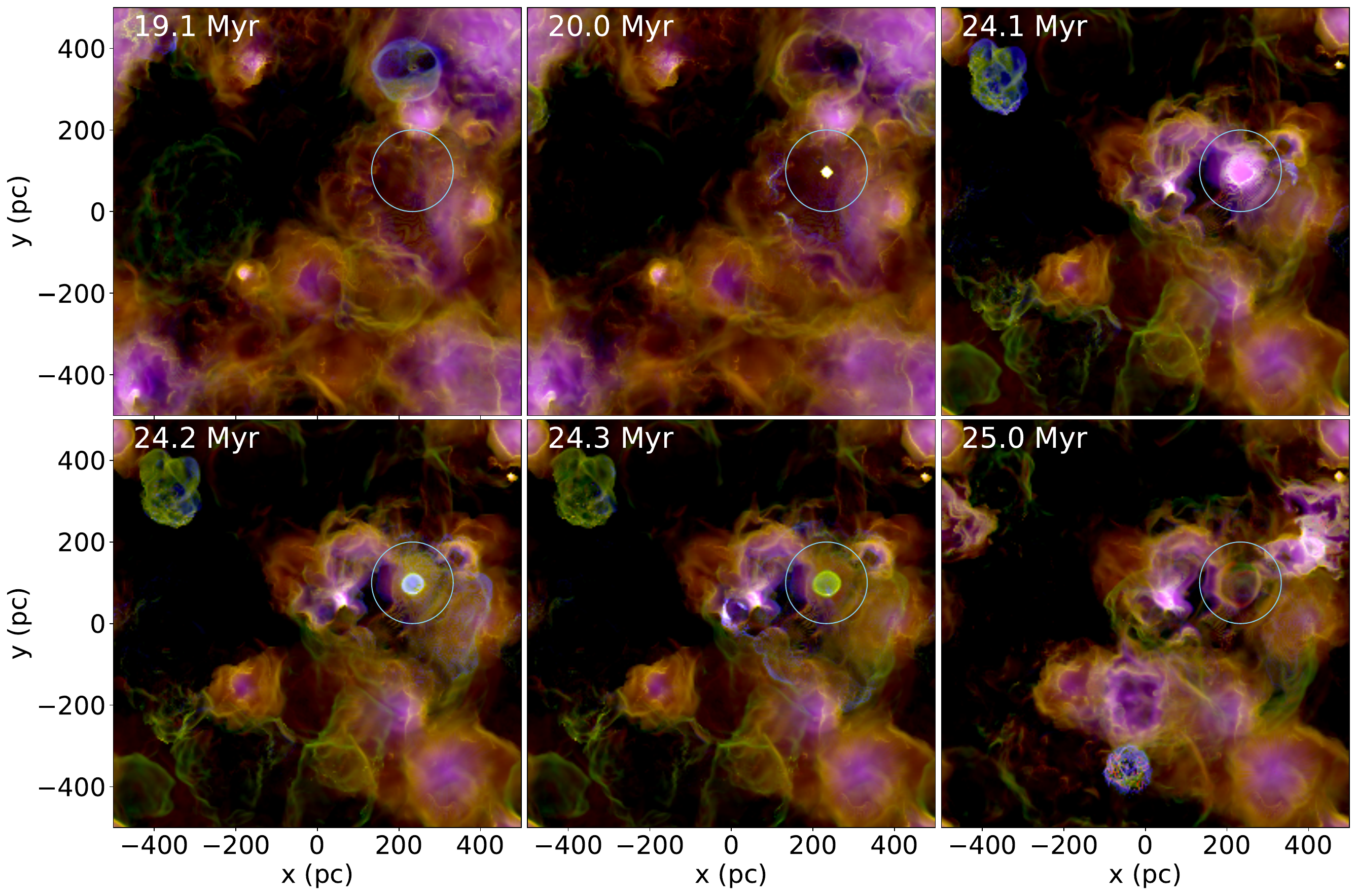}
   \caption{Evolution in the vicinity of a constrained `$\zeta$ Ophiuchi-like' massive star throughout it's lifetime. Snapshots have been selected to show key moments in the evolution. Reading from top left to bottom right the snapshots show; the ISM before the star is born (19.1~Myr), the H{\sc{ii}} region formed soon after the birth of the star (20.0~Myr), the ISM having been disrupted by the ionizing radiation from the star, creating a more spread out and diffuse H{\sc ii} region (24.1~Myr), the ISM very soon after the star goes supernova, shining brightly in a shell of blue [O{\sc iii}] (24.2~Myr), the supernova shock having cooled and recombined into a state which shines brightly in the lower energy ion of [S{\sc ii}] (24.3~Myr), the supernova shock beginning to dissipate into the wider turbulent ISM (25.0~Myr).}
   \label{moviesnaps}  
\end{figure*}

Having identified these 3 different O-star environments, we now look to our \emph{time dependent} radiation-hydrodynamics simulation for time-dependent examples of similar objects. Figure~\ref{moviesnaps} shows 6 snapshots of the evolution of a `$\zeta$ Ophiuchi-like' O-star. In appendix~\ref{appendix2}, figure~\ref{moviesnaps1} shows similar snapshots for a `Bajamar-like' case, and figure~\ref{moviesnaps2} shows the evolution of a `$\zeta$ Puppis-like' O-star.

Figure~\ref{moviesnaps} displays the near spherically symmetrical evolution of our `$\zeta$ Ophiuchi-like' star. Once the star is born, the ionization front is seen to evolve over a period of around 100~kyr, leading to a classical H{\sc ii} region which is nearly spherical. It shines brightly and compactly in all of H$\alpha$, [O{\sc iii}], and from the edges of the ionized region, [S{\sc ii}], as the dynamical effect of the heating by LyC photons works to disrupt the dense cloud within which it was born. Over the star's lifetime of around 4~Myr, it disperses its birth cloud into a larger, more diffuse structure with a more dense spherical shell surrounding it. The inside of this structure shines brightly in [O{\sc iii}] and H$\alpha$, and [S{\sc ii}] is confined mostly to the dense ring, again at the edge of the ionized region. When the star goes supernova, a blue [O{\sc iii}] shell is formed by collisional ionization which persists for around 100~kyr. H$\alpha$ is suppressed in this phase due to the high temperature of the supernova remnant inhibiting proton/electron recombinations. After the `blue-bright' phase, the remnant has cooled and the oxygen has recombined to lower ionization states than O{\sc iii}. The hydrogen gas is also neutral, but the temperature of the gas remains in the $10^{4}~\rm K$ regime. This creates a shell which shines brightly in [S{\sc ii}] (with a lower ionization potential than hydrogen) without emitting in H$\alpha$. This phase appears as green in the RGB images, and would represent a SNR which is detected at high values of [S{\sc ii}]/H$\alpha$, consistent with what is seen in \citet{caldwell25} (SNR having [S{\sc ii}]/H$\alpha$ > 0.4). 

Figure~\ref{moviesnaps1} (in appendix~\ref{appendix2}) shows the evolution of our `Bajamar-like' star. This star evolves in a very similar way to the `$\zeta$ Ophiuchi-like' example, but with a low density channel allowing ionization by LyC photons in the positive $x$ direction. This channel doesn't appear to morphologically affect the way in which the resulting SNR evolves, with both remnants evolving in a near spherical shell. Whether this is simply a feature of the subgrid supernova models is yet to be explored. Note that there is no `blue phase' of the SNR evolution for this example. Due to the higher density of the environment in which the supernova occurs, the recombination of higher energy ions happens over a shorter timescale, and the `blue phase' is not resolved in the 100~kyr temporal resolution of the figure.

For the `$\zeta$ Puppis-like' example shown in figure~\ref{moviesnaps2} (appendix~\ref{appendix2}), the massive star is formed in a relatively low density region due to the star formation algorithm used in the simulations. While not necessarily physical, these formation events are included to represent stars which either move quickly from their birth clouds to low density regions or have their birth cloud disrupted by other stellar feedback. O-stars in these environments clearly exist in nature (demonstrated in figure~\ref{boundedness} for $\zeta$ Puppis). When this star turns on, a very large ionized region quickly develops over a distance of some 400~pc, illuminating folds and filaments which were previously unseen in the neutral gas. The low density gas in this enormous H{\sc ii} region shines brightly in [O{\sc iii}] and H$\alpha$, but is too highly ionized to shine in [S{\sc ii}].

Over a few Myr, the folds and filaments are seen to be dynamically pushed away from the luminous ionizing source, and appear to be `eroded' away, leaving tadpole-shaped, round-headed objects with long tails shielded from the photoionization feedback. These structures bear resemblance to the so called `Pillars of Creation' which have been observed in the Eagle nebula \citep{hillenbrand93,mcleod15}. These structures are also noted in the simulations of \citet{ali18}. Note that these simulations do not include stellar winds, and this shaping effect is purely due to photoionization feedback. The supernova event follows a similar evolution to the other two examples, but being a lower density environment, cools and recombines over a slightly longer timescale than the $\zeta$ Ophiuchi and Bajamar example.

\section{Discussion} \label{discussion_section}

This paper is the fourth in a series investigating the nature of the warm ionized medium of the Milky Way in the vicinity of the Sun.  In the first \citep{mccallum24a}, we used hydrodynamical simulations combined with Monte Carlo radiative transfer and photoionization to explore what was needed to match the {\it observed kiloparsec density scaleheight} of ionized gas, finding that we required a star formation rate of 1200~$\rm M_{\odot}~\rm{yr}^{-1}~\rm{kpc}^{-2}$ and demonstrating that the inclusion of non-equilibrium ionization was an important effect to include as resulted in persistent ionized gas due to the long recombination times of the low density DIG. The second paper \citep{mccallum24b} showed that non-equilibrium ionization was also an important effect to include in order to predict the {\it observed intensity of the metal lines as a function of height},  particularly for species like Ne{\sc iii}.  In the third paper \citep{mccallum25}, we took advantage of new 3D dust maps to use the observationally constrained density structure of the ISM as the input for radiative transfer models, rather than the density structure from hydrodynamical models, and found a remarkably good match to the observed overall intensity and morphology of the observed H$\alpha$ sky.  

In this work, the normalisation of the Kennicutt-Schmidt relation has been tuned in order to morphologically match the structure of the local Milky Way. This is in contrast to our previous work in which the SFR was set in order to support a vertical layer of ionized gas which matches Galaxy-wide observational estimates for the scale heights of diffuse ionized. It is notable that the star formation rates for these two purposes are not the same. For the former, a value of 370~$\rm M_{\odot}~\rm{Myr}^{-1}~\rm{kpc}^{-2}$ was found to provide the closest match the structure of the Milky Way (both seen in the dust maps, and the used population of O stars), whereas 1200~$\rm M_{\odot}~\rm{yr}^{-1}~\rm{kpc}^{-2}$ was required to produce DIG with a kpc scale height. The scale height of the DIG layer in the 370~$\rm M_{\odot}~\rm{Myr}^{-1}~\rm{kpc}^{-2}$ run before truncation of the simulation box was found to be 400~pc. One possibility is that that the midplane region of the Milky Way (out to approximately 500~pc in each direction of the $x-y$ plane) is more quiescent now and the higher altitude gas is not in dynamic equilibrium, but is relaxing after a period of higher star formation. This would be consistent with the result from \cite{swiggum24} that a burst of star formation was initiated in the solar neighborhood approximately 45 Myr ago and has been declining to the present era. This extreme variation in SFR is not currently seen our simulations due to our adopted prescription for star formation, however future work could involve enforcing an observationally motivated star formation history in order to produce a 1~kpc scale height of WIM, whilst simultaneously matching the relatively quiescent state of our local region. 

Although the high rate of star formation that was required in our first paper may not be consistent with the nearby morphological constraints, a comparison of our higher star formation rates are similar to the images of the Small Magellanic Clouds (figure~\ref{smc}) and might well have been appropriate to the solar neighborhood $\sim 30-45$ Myr ago. 

In comparing our \emph{photoionization-only} models and \emph{time dependent} models, we find that the H$\alpha$ emissivity grids are in close agreement. There are three reasons for this.  First, at the higher densities present in the midplane, the hydrogen recombination times are shorter and departures from ionization equilibrium are less significant than for the low density gas at high altitude. Recombination timescales in $1~\rm cm^{-3}$ gas are around 0.1~Myr versus 10~Myr in $0.01~\rm cm^{-3}$ gas. Second, the regions of the simulation that are affected by supernova shocks are at higher temperature (which lowers the H$\alpha$ emissivity per H atom) and lower density, which reduces the observed emission measure. All of these effects tend to diminish the role of time-dependent ionization and shock heating in affecting the observed H$\alpha$ emission. This may be one of the reasons why the simulations of the H$\alpha$ sky reported in \citet{mccallum25} were in such good morphological agreement with observations despite the assumption of photoionization equilibrium and the absence of shocks or supernova remnants. 

\begin{figure}
    \centering
    \includegraphics[width=0.75\linewidth]{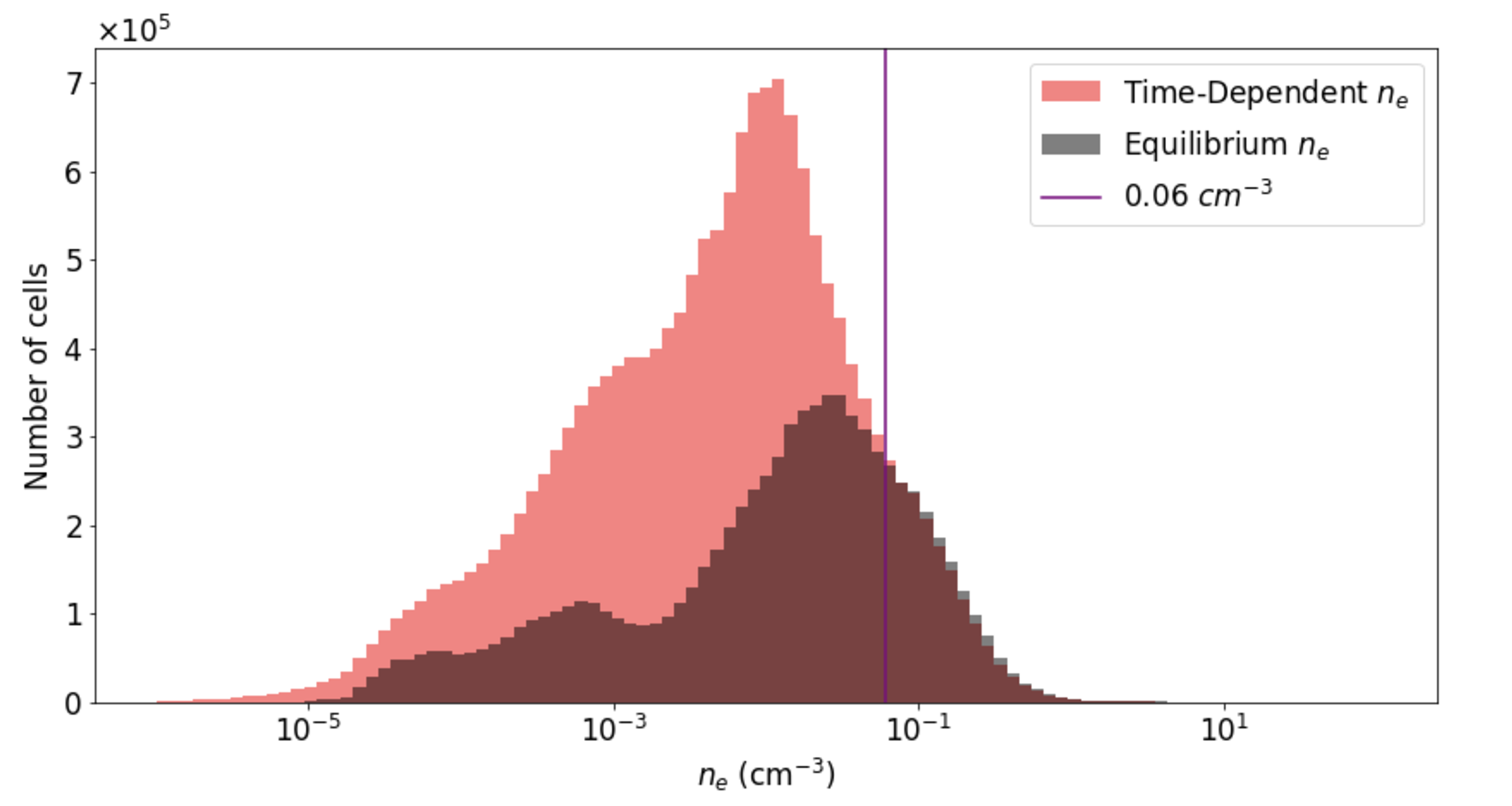}
    \caption{Histogram of electron densities in both the \emph{time dependent} snapshot, and the \emph{photoionization-only} snapshot of the LOW SFR run. The histograms are seen to diverge at densities below 0.06~$\rm cm^{-3}$. There are fewer ionized cells in the equilibrium histogram, as the equilibrium run has a number of cells with zero ionized fraction, leading to cells which don't contribute towards this histogram. Note the histograms match above electron densities of 0.06~$\rm cm^{-3}$.}
    \label{fig:denscutoff}
\end{figure}

This is confirmed by looking at the histogram of electron densities in the \emph{time dependent} and \emph{photoionization-only} models of the 370~$\rm M_{\odot}~\rm{Myr}^{-1}~\rm{kpc}^{-2}$ simulation. In both cases, the histograms match for densities above 0.06~$\rm cm^{-3}$ (see figure~\ref{fig:denscutoff}). We can thus expect to recover all ionized gas above this threshold in our \emph{observational} simulation. Around 76\% of the ionized mass is contained in cells above this density. We find that the \emph{photoionization-only} run contains 93\% of the total ionized mass of the \emph{time dependent} simulation.

In contrast, the predictions of the other optical emission lines, particularly [S{\sc ii}] 6716{\AA}, [N{\sc ii}] 6584{\AA}, [O{\sc iii}] 5007{\AA} and [Ne{\sc iii}] 15$~\mu m$, are  more sensitive to the presence of collisional ionization and thermal heating due to the effect of supernova blast waves. In trying to match observations of these emission lines to the sky, it is more important to consider these time-dependent effects. This also provides a way to search for areas of recent energy input into the ISM, complementing previous efforts to find regions of supernova energy input using X-ray and radio synchrotron emission and expanding HI shells.  

Finally, these simulations may be used to study the diversity of outcomes of stellar births and supernova events as a function of the interstellar environment. For each massive star formed in these simulations we tend to see a similar general evolution (see figure~\ref{moviesnaps}): bright H$\alpha$ and [O{\sc iii}] emission, an expanding high excitation H{\sc ii} region, a bright and fast expanding [O{\sc iii}] ring following the supernova detonation, and then a slower, fainter ring of [S{\sc ii}] emission that merges back into the general ISM in about $\sim$2 Myr. While this general evolution is consistent from star to star, the volume of ionized gas generated by any one star can vary enormously depending on stellar environment. We find that while the Bajamar star and $\zeta$ Puppis have similar ionizing luminosities, the LyC photons from the latter are spread out over a volume which is 827 times the former, suggesting that the DIG may be formed by a small number of very luminous O-stars which exist in environments which allow their LyC photons to escape their birth clouds and travel hundreds of parsecs.

Continued multi-wavelength optical line emission observations of the sky should in the future provide new insights into the recent history of the energy input into the local interstellar medium.

\section{Conclusions} \label{conclusion_section}

We have presented the results of our 3D simulations of the photoionized gas in the 1.25~kpc sphere around the sun. We have shown synthetic skies for a series of collisionally excited optical and infrared emission lines, many of which have not yet been observed in large portions of the sky, positioning these simulations as predictions for future surveys.

Through a suite of zoom-in radiation-hydrodynamics simulations, we have constrained the local Milky Way SFR (over a 1~$\rm kpc^{2}$ patch) to be a factor of 4 lower than the value required to reproduce Galaxy-wide observations of DIG scale heights (370 versus 1200~$\rm M_{\odot}~Myr^{-1}~kpc^{-2}$). This lower SFR is consistent with the Sun's location in a relatively quiescent region of the Galaxy. The \emph{photoionization-only} snapshots allowed us to qualitatively assess the limitations of the static \emph{observational} simulation, identifying the primary missing ingredient as supernova-driven structures, including ionized shells and bubbles.

The \emph{time dependent} simulations show that supernova-heated, non-equilibrium gas can populate regions of diagnostic diagrams often associated with LI(N)ER emission, even in the absence of evolved stellar populations. This reinforces the idea that non-equilibrium processes and local ISM dynamics alone may explain some observed LI(N)ER-like features in the Milky Way and other galaxies.

The impact of missing supernovae on the predicted H$\alpha$ emission is seen to be minimal due to the inverse temperature dependence of the H$\alpha$ emissivity, which suppresses the contribution of hot, shock-heated gas to the total H$\alpha$ intensity. This supports the reliability of the H$\alpha$ skies presented in \citet{mccallum25}.

We have used our radiation-hydrodynamics simulations to illustrate the general evolution of emission from H{\sc ii} regions and SNRs. We highlight three examples of varying O-star environment, and identify similar examples within our best fitting radiation-hydrodynamics simulation. The effect of O-star environment is seen to be crucial in any individual star's ability to ionize a large volume of diffuse ionized gas. We see that SNRs in our simulated ISM follow an evolution from being bright in [O{\sc iii}] (for around 100-200~kyr) to bright in [S{\sc ii}] (for around 1-2~Myr). With these timescale estimates we predict a number ratio of [O{\sc iii}]-bright to [S{\sc ii}]-bright SNRs as approximately 1:10.

The simulations presented in this paper provide not only morphological predictions but also diagnostic line ratio distributions that will be valuable for interpreting forthcoming optical and infrared emission line surveys. Future work incorporating radiative shocks, X-ray photoionization, and more detailed stellar populations will be essential to fully capture the evolution and emission properties of the ionized ISM.

\section*{Acknowledgements}

LM acknowledges financial support from a UK-STFC PhD studentship. DK acknowledges support from HST-AR-17060 provided by NASA through a grant from the Space Telescope Science Institute, which is operated by the Association of 389 Universities for Research in Astronomy, Inc., under NASA 390 contract NAS5-26555.

\section*{Data Availability}

3D cubes of selected line emissivities from the high resolution \emph{observational} model (out to 1.25~kpc) are available for download at \url{https://doi.org/10.5281/zenodo.15710752}. Further emission lines or grids of ionic fractions can be made available on request.



\bibliographystyle{mnras}
\bibliography{biblio} 




\appendix

\section{Heating and Cooling - Static Equilibrium Case}
\label{appendix1}

For the static equilibrium case of the \emph{observational} and \emph{photoionization only} simulations, the gas temperature in each cell is fixed by the condition $G(T)=L(T)$, with \(G\) the total heating rate per unit volume and \(L\) the total cooling rate.

\subsection{Photoionization Heating}

We use only photoionization of hydrogen and helium as a heating
source.  For neutral helium the heating rate is
\begin{equation}
G(\mathrm{He}^{0}) = n(\mathrm{He}^{0})
\int_{\nu_{\mathrm{He}^{0}}}^{\infty}
\frac{4\pi J_\nu}{h\nu}\,
a_\nu(\mathrm{He}^{0})\,h(\nu - \nu_{\mathrm{He}^{0}})\,
\mathrm{d}\nu
\end{equation}
and the corresponding Monte-Carlo integral is
\begin{equation}
I_{H}^{\mathrm{He}} =
\int_{\nu_{\mathrm{He}^{0}}}^{\infty}
\frac{4\pi J_\nu}{h\nu}\,
a_\nu(\mathrm{He}^{0})\,h(\nu - \nu_{\mathrm{He}^{0}})\,
\mathrm{d}\nu
=
\Bigl[\,\tfrac{Q}{N\,V}\Bigr]\;
l\,a_\nu(\mathrm{He}^{0})\,
\bigl(h\nu - h\nu_{\mathrm{He}^{0}}\bigr). 
\end{equation}

Under the on-the-spot approximation, He \textsc{i} Ly$\alpha$
re-processing contributes an extra term  
\(P(\mathrm{H_{OTS}})\,n(\mathrm{He}^{+})\,n_e\,
\alpha_{\!2\,^1\!P}^{\mathrm{eff}}(\mathrm{Ly}\alpha)
\,[h\nu(\mathrm{Ly}\alpha)-\chi(\mathrm{H}^{0})]\)  
to the heating budget of each cell.

\subsection{Cooling by Recombination and Free–Free Emission}

Recombination cooling curves for \(\mathrm{H}^{+}\) and
\(\mathrm{He}^{+}\) are taken from \citet{hummer94} and
\citet{hummer98}.  The free–free contribution is
\begin{equation}
L_{\mathrm{ff}} = 1.42\times10^{-27}
\bigl[n(\mathrm{H}^{+})+n(\mathrm{He}^{+})\bigr]\,
n_e\,g_{\mathrm{ff}}\,
\sqrt{T_e}
\end{equation}
with Gaunt factor
\begin{equation}
g_{\mathrm{ff}} = 1.1 + 0.34
\exp\!\Bigl[-(5.5-\log T_e)^{2/3}\Bigr]
\end{equation}
fitted by \citet{katz96} to the tabulation of \citet{spitzer78}.

\subsection{Cooling from Forbidden-Line Excitation}

In addition, we include cooling contributions from the collisionally excited optical and infrared emission lines, calculated as described in \citet{cmacionize}.


\section{Evolution of Individual Sources}
\label{appendix2}

Figures showing the evolution of our `Bajamar-like' case and our `$\zeta$ Puppis-like' case star.

\begin{figure*}
    \centering
    \includegraphics[width=1.0\textwidth]{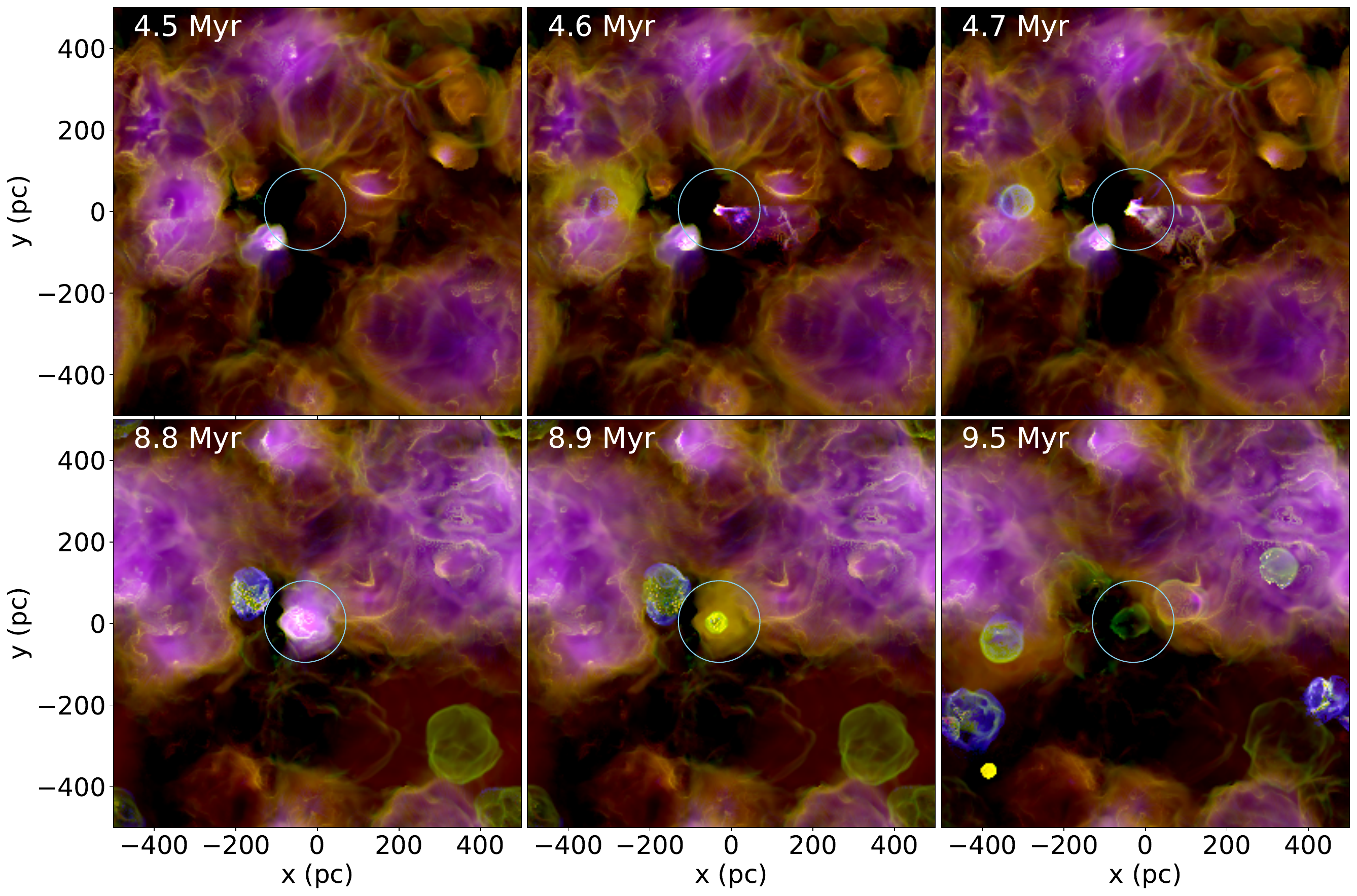}
    \caption{Typical evolution in the vicinity of a semi-constrained `Bajamar-like' massive star throughout it's lifetime. Snapshots have been selected to show key moments in the evolution. Reading from top left to bottom right the snapshots show; the ISM before the star is born (4.5~Myr), the H{\sc{ii}} region formed soon after the birth of the star, showing a cone of ionization due to the escape of LyC photons  in the positive X direction (4.6~Myr), the ionized cone having become brighter and more ionized (4.6~Myr), the ISM having been disrupted by the ionizing radiation from the star, creating a more spread out and diffuse H{\sc ii} region (8.8~Myr), the ISM very soon after the star goes supernova, shining brightly in a shell of green [S{\sc ii}] (8.9~Myr), the supernova shock beginning to dissipate into the wider turbulent ISM (9.5~Myr).}
    \label{moviesnaps1}
    
\end{figure*}

\begin{figure*}
   \centering
   \includegraphics[width=1.0\textwidth]{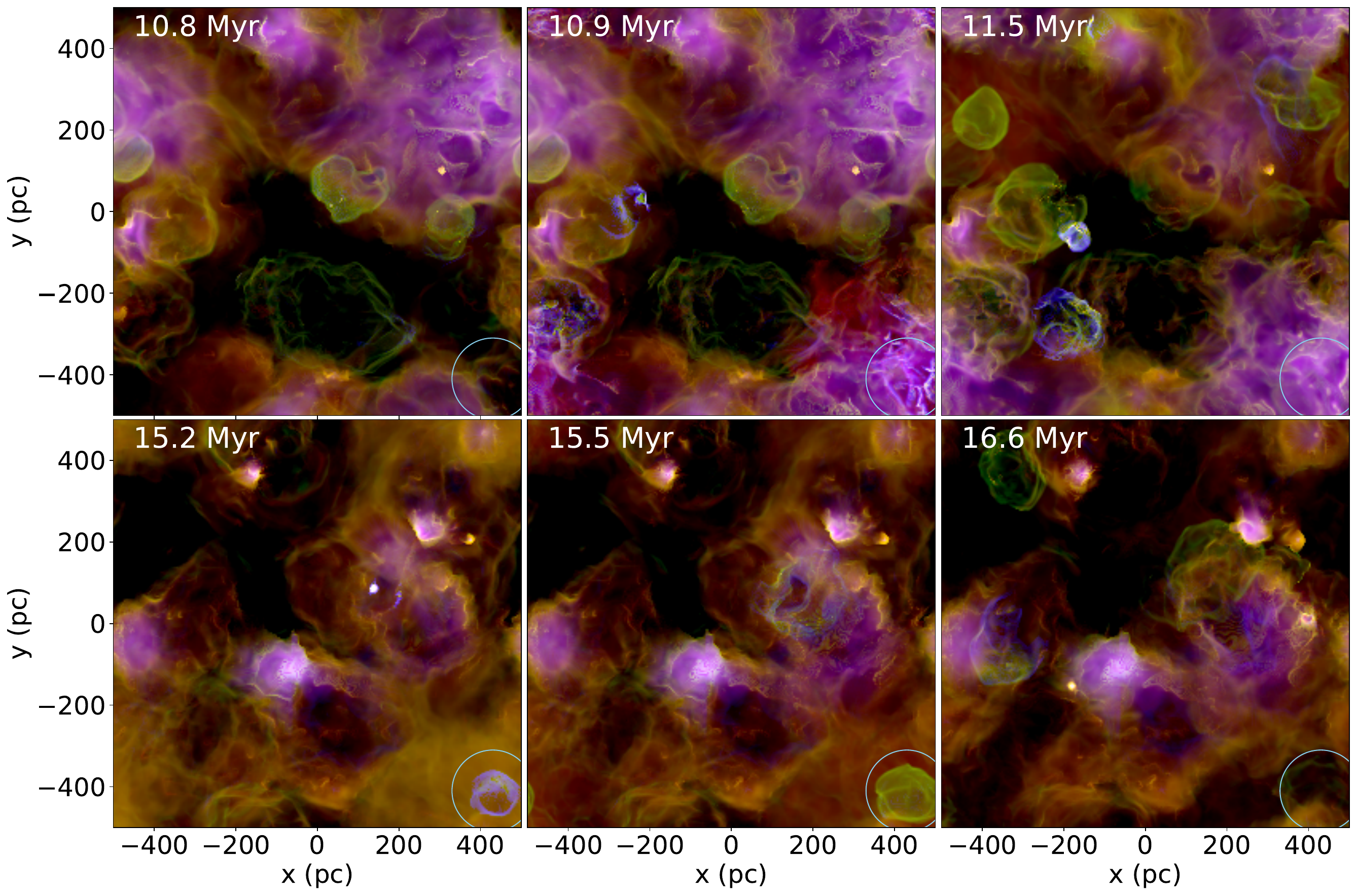}
   \caption{Typical evolution in the vicinity of a `$\zeta$ Puppis-like' star throughout it's lifetime, whereby the star exists in a low density environment. Snapshots have been selected to show key moments in the evolution. Reading from top left to bottom right the snapshots show; the ISM before the star is born (10.8~Myr), the large and diffuse H{\sc{ii}} region formed soon after the birth of the star (10.9~Myr), the ISM having been disrupted by the ionizing radiation from the star, pushing material into eroding columns much like the `Pillars of Creation' (11.5~Myr), the ISM very soon after the star goes supernova, shining brightly in a shell of blue [O{\sc iii}] (15.2~Myr), the supernova shock having cooled and recombined into a state which shines brightly in the lower energy ion of [S{\sc ii}] (15.5~Myr), the supernova shock beginning to dissipate into the wider turbulent ISM (16.6~Myr).}
   \label{moviesnaps2}    
\end{figure*}


\bsp	
\label{lastpage}
\end{document}